%% file: main.tex
\documentclass[11pt, logo, onecolumn, copyright, colorlinks=true, allcolors=blue]{nvidiatechreport}

\usepackage[english]{babel}

\usepackage{amsmath}
\usepackage{multirow}
\usepackage{graphicx}
\usepackage[section]{placeins}
\usepackage{flafter}
\usepackage{listings}
\usepackage{cite}

\usepackage[colorlinks=true, allcolors=blue]{hyperref}

\makeatletter
\AtBeginDocument{%
    \def\ttl@Hy@refstepcounter#1{%
        \LinkTargetOff
        \refstepcounter{#1}%
        \LinkTargetOn
        \def\ttl@Hy@saveanchor{%
            \AddToHookNext{para/begin}{\MakeLinkTarget{#1}}%
        }%
    }%
}
\makeatother

\title{Generalized Tensor Parallelism: Powering Trillion-Parameter LLM Training at 10+ EFLOP/s}
\author{Shiqing~Fan\textsuperscript{\textdagger},
Deepak~Narayanan\textsuperscript{\textdagger}, Jimmy~Zhang, Jiangfei~Duan,
Prajwal~Singhania, Haoran~Zhang, Wanying~Wang, Philip~Monk, Xuwen~Chen, Liangyu~Zhao,
Jared~Casper, Jun~Yang, Mohammad~Shoeybi, Bryan~Catanzaro,
and Vijay~Anand~Korthikanti\thanks{Leads the effort.}\textsuperscript{,\textdagger}\par\vspace{4pt}
NVIDIA}

\begin{document}

\begin{abstract}
\textbf{Abstract.} Pretraining state-of-the-art large language models (LLMs) with up to trillions of parameters is memory- and compute-intensive.
Existing parallelization schemes compromise on either per-GPU memory or throughput: data parallelism replicates parameters; tensor parallelism and fully sharded data parallelism (FSDP) add communication of activations and parameters, respectively, to every microbatch; and pipeline parallelism leaves GPUs idle while the pipeline fills and drains.
We show that, for linear layers, five commonly used parallelisms (data, tensor, sequence, context parallelism, and FSDP) are special cases of one framework, generalized tensor parallelism (GTP), in which each layer's parameters and activations are partitioned and either can be re-materialized with an all-gather before use.
In GTP, a parameter all-gather is an optional preprocessing step on one input of a matrix multiplication (GEMM), whereas FSDP implementations typically treat it as a step before an entire Transformer block, which can increase memory footprint and delay GEMMs.
The GTP view lets us schedule all-gathers per GEMM rather than per block, issuing each early enough to hide its communication but late enough to limit memory.
Our implementation is open source at \url{https://github.com/NVIDIA/Megatron-LM}; with it, an internal 2T-parameter mixture-of-experts (MoE) model reaches \textbf{13.0 model EFLOP/s} (exaFLOP/s) on 12,288 GB300 GPUs with 98\% weak-scaling efficiency, and runs with 1M-token sequences on 512 GPUs.
On 256--1,024 GPUs, our implementation achieves \textbf{1.05--1.55$\times$ the throughput of Megatron-FSDP} across three MoE models and \textbf{1.27--1.31$\times$ that of a tuned 3D-parallel recipe with pipeline parallelism} on a DeepSeek-V4-Pro proxy model.
\end{abstract}

\begingroup
\renewcommand{\thefootnote}{\fnsymbol{footnote}}
\maketitle
\makeatletter
\@thanks
\makeatother
\footnotetext[2]{Correspondence to: Shiqing Fan
\textless\href{mailto:shiqingf@nvidia.com}{\nolinkurl{shiqingf@nvidia.com}}\textgreater,
Deepak Narayanan
\textless\href{mailto:dnarayanan@nvidia.com}{\nolinkurl{dnarayanan@nvidia.com}}\textgreater,
and Vijay Anand Korthikanti
\textless\href{mailto:vkorthikanti@nvidia.com}{\nolinkurl{vkorthikanti@nvidia.com}}\textgreater.}
\endgroup
\setcounter{footnote}{0}

\input{sections/1_introduction}

\input{sections/2_gtp}

\input{sections/3_parameter_scheduling}

\input{sections/4_implementation}

\input{sections/5_evaluation}

\input{sections/6_parallelism_tradeoffs}

\input{sections/7_related_work}

\input{sections/8_conclusion}

\input{sections/9_acknowledgments}

\bibliography{main}
\bibliographystyle{plain}

\end{document}

%% file: sections/1_introduction.tex
\section{Introduction}
\label{sec:intro}

Modern large language models (LLMs) have up to trillions of parameters~\cite{2020gpt3,2023palm,2024llama3,2024deepseekv3,2025llama4,2026deepseekv4}.
The parameters, optimizer state, and activations of these models far exceed the memory of a single GPU.\footnote{Our arguments also apply to other accelerators, such as TPUs, that have a fast interconnect within a group of devices and a slower network between groups.}
Training these models on trillions of tokens requires high throughput to keep training time manageable~\cite{2022chinchilla,2024llama3,2024deepseekv3}.
Efficient distributed pretraining over thousands of GPUs thus has two key requirements.
First, per-GPU memory for parameters, optimizer state, and activations should shrink roughly in proportion to GPU count, so that larger models and longer sequences fit in the available memory capacity.
Second, GPUs should stay busy: communication should overlap with computation, and per-GPU matrix multiplications (GEMMs) should run efficiently.

Existing parallelization schemes each compromise on at least one of these requirements.
Data parallelism (DP) replicates parameters on every GPU, so parameter memory does not shrink as GPUs are added; a distributed optimizer~\cite{2020zero} can shard optimizer state.
Tensor parallelism (TP)~\cite{2019megatron} shards parameters and can be combined with sequence parallelism (SP)~\cite{2023activationrecomputation} to additionally shard the activations between tensor-parallel regions.
However, TP communicates activations in every layer for every microbatch.
This communication is hard to hide and is fast only within an NVLink domain, which bounds the TP degree; large TP degrees also shrink GEMMs, decreasing arithmetic intensity and GPU utilization.
Pipeline parallelism (PP)~\cite{2019gpipe,2019pipedream,2021megatronscaling} communicates little, but naively retaining activations for multiple in-flight microbatches leads to an activation footprint similar to that of undistributed computation; it also leaves GPUs idle while the pipeline fills and drains and requires compute to be relatively balanced across stages.
Fully sharded data parallelism (FSDP) and ZeRO-3~\cite{2020zero,2023pytorchfsdp} shard parameters, gradients, and optimizer state while keeping GEMMs large (given a large enough batch), but all-gather every parameter for every microbatch in both the forward and backward passes, adding communication.
Context parallelism (CP)~\cite{2023sequenceparallelism,2023deepspeedulysses,2023ringattention,2025torchtitancp} splits each sequence across GPUs to reduce activation memory.
Expert parallelism (EP)~\cite{2021gshard,2026mcoremoetraining} distributes the experts of mixture-of-experts (MoE) models across GPUs; both CP and EP compose with the schemes above.

In this paper, we show that, for a linear layer, five commonly used parallelisms (DP, TP, SP, CP, and FSDP) are special cases of a single framework, which we call generalized tensor parallelism (GTP; Section~\ref{sec:gtp}).
In GTP, parameter matrices are partitioned along a hidden dimension, as in TP, and input activations along the token dimension.
Token partitions can contain whole sequences (DP) or parts of each sequence (CP).
Linear layers process these partitions independently; with CP, attention and state-space layers also exchange their state (e.g., keys and values for attention layers) among the GPUs that hold parts of each sequence.
Each linear layer's GEMM thus has two sharded inputs, parameters and activations, and either can be re-materialized with an all-gather before use: activations for SP, or parameters for FSDP.
The partitioning degrees and the number of GPUs across which each input is re-materialized trade communication against memory, and can be chosen separately for dense and expert weights.
When both inputs are fully sharded, per-GPU memory for parameters, optimizer state, and activations shrinks in proportion to the number of GPUs.

Parameter re-materialization reduces memory but is not a free lunch: each microbatch all-gathers the parameters in both the forward and backward passes and reduce-scatters their gradients ($1.5\times$ the volume of DP's once-per-step gradient synchronization at equal communication precision; Section~\ref{subsec:gtp-communication-analysis}).
Unlike TP's activation communication, a parameter all-gather does not depend on the preceding computation, so it can be issued ahead of the layer that consumes it.
When to issue each all-gather is thus a choice with consequences: re-materializing too close to the compute that needs the parameters exposes communication time~\cite{2023pytorchfsdp,2026vescalefsdp}, while re-materializing too far ahead increases memory footprint.
Existing FSDP implementations schedule re-materialization in configurable units, often Transformer blocks~\cite{2017transformer} (Sections~\ref{subsec:gtp-fsdp-problems} and~\ref{sec:related}).
When the unit is a whole block, its first GEMM waits for all of the block's parameter all-gathers, and all of the block's parameters are in memory at once.
In GTP, a parameter all-gather is instead the counterpart of SP's activation all-gather: a preprocessing step on one input of a GEMM.
Scheduling parameter re-materialization per GEMM rather than per block lets us tailor when each all-gather is issued, what each GEMM waits for, and which parameters share a collective, based on computation and memory requirements.
For MoE layers, we use this flexibility to issue each expert weight's all-gather a full MoE block ahead of its GEMM (Section~\ref{subsec:gtp-scheduling-tradeoffs}).

To summarize, we make the following contributions:
\begin{itemize}[leftmargin=*, itemsep=2pt, topsep=4pt]
\item \textbf{A unified framework.} We express DP, TP, SP, CP, and FSDP as special cases of parameter and activation partitioning with optional input re-materialization (Section~\ref{sec:gtp}).
\item \textbf{Per-GEMM re-materialization scheduling.} We schedule parameter all-gathers per GEMM through separate prefetch chains (C1), consumer-specific waits (C2), and coalesced expert all-gathers (C3; Section~\ref{subsec:gtp-scheduling-tradeoffs}); on two benchmark models using 256 GPUs, our full implementation is 4.6--13.9\%, 2.8--18.7\%, and 3.5--9.0\% faster than variants without C1, C2, and C3, respectively (Section~\ref{subsec:gtp-eval-ablations}).
\item \textbf{An open implementation, evaluated at scale.} Our GTP implementation is open source in Megatron-Core (MCore) and supports dense and MoE models with AdamW and Muon; with it, an internal 2T-parameter MoE model (Internal-2T) reaches 13.0 model EFLOP/s at a sequence length of 8K on 12,288 GB300 GPUs, with 98\% weak-scaling efficiency from 4,096 to 12,288 GPUs (Section~\ref{subsec:gtp-eval-scaling}).
GTP also enables the full Internal-2T model to run with sequences up to 1M tokens on 512 GPUs (Section~\ref{subsec:gtp-eval-long-sequence}).
Across three MoE models at 256--1,024 GPUs, our implementation achieves 1.05--1.55$\times$ the throughput of Megatron-FSDP (MFSDP; Section~\ref{subsec:gtp-eval-fsdp2}).
On a DeepSeek-V4-Pro (DSV4-Pro) proxy model, it achieves 1.27--1.31$\times$ the throughput of a tuned 3D-parallel recipe with pipeline parallelism (Section~\ref{subsec:gtp-eval-3d}).
The code is available at \url{https://github.com/NVIDIA/Megatron-LM}.
\end{itemize}

\paragraph{Organization.}
Section~\ref{sec:gtp} describes the GTP framework and analyzes its communication and memory.
Section~\ref{sec:design} explains how we schedule parameter all-gathers per GEMM to hide communication and limit memory.
Section~\ref{sec:implementation} describes our implementation in MCore and Transformer Engine.
Section~\ref{sec:evaluation} empirically compares GTP with appropriate baselines.
Section~\ref{sec:design-space} offers practical guidance for configuring GTP, including when to combine it with PP.
Section~\ref{sec:related} reviews related work.

%% file: sections/2_gtp.tex
\section{Generalized Tensor Parallelism}
\label{sec:gtp}

Most of an LLM's parameters are in linear layers: attention projections, MLPs, and the experts of MoE layers.
Each linear layer is a GEMM between two inputs: an activation matrix $X$, with one row per token, and a parameter matrix $W$.
Distributing these layers comes down to deciding how to split and store $X$ and $W$ across GPUs.
In this section, we show that DP, TP, SP, CP, and FSDP each make different choices for $X$ and $W$, and that all five are special cases of a single framework, generalized tensor parallelism (GTP; Section~\ref{subsec:gtp-framework}).
We then analyze the communication (Section~\ref{subsec:gtp-communication-analysis}) and memory (Section~\ref{subsec:gtp-memory}) costs of these choices.
PP~\cite{2019gpipe,2019pipedream,2021megatronscaling} is orthogonal: it partitions the model's layers into stages that exchange only activations and their gradients point-to-point.
GTP composes with PP; Section~\ref{subsec:gtp-pp-tradeoffs} discusses when to combine them.

Other layers need slightly different treatment.
Small weight tensors, such as LayerNorm~\cite{2016layernorm} or RMSNorm~\cite{2019rmsnorm} weights, can be replicated to reduce needless communication.
The attention operation itself has no parameters, and its communication under CP is the same with or without GTP.

\subsection{One Linear Layer, Two Inputs}
\label{subsec:gtp-framework}

\begin{figure}[!htbp]
    \centering
    \includegraphics[width=0.5\columnwidth]{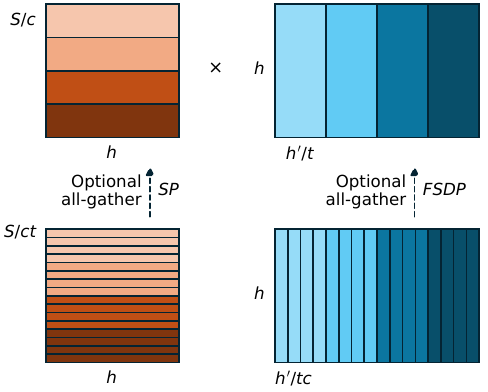}
    \caption{Partitioning and re-materialization of a GEMM with $t=c=4$.
    Top: parameters are partitioned across $t$ GPUs and activation rows (tokens) across $c$ GPUs; each shade is one partition.
    Bottom: each input can optionally be stored as a smaller shard and all-gathered before the GEMM, activations across the $t$-group (SP) and parameters across the $c$-group (FSDP).}
    \label{fig:gtp-full-sharding}
\end{figure}

As a running example, we use an MLP block: an $h \rightarrow h'$ GEMM followed by an $h' \rightarrow h$ GEMM, where $h$ is the model's hidden size and $h'$ the MLP's intermediate size (e.g., $h' = 4h$ in many dense models); other linear layers behave similarly.

TP~\cite{2019megatron} partitions the parameters along a hidden dimension across $t$ GPUs: the $h \rightarrow h'$ GEMM is split by output columns and the $h' \rightarrow h$ GEMM by input rows (column- and row-parallel layers, respectively).
Each GPU then computes a slice of every layer, and the GPUs combine their partial outputs with an all-reduce at the end of each tensor-parallel region.
The all-reduce depends on the preceding GEMM's output, so overlapping the two requires splitting the GEMM into smaller GEMMs~\cite{2023overlapdecomposition,2024domino}, which further lowers arithmetic intensity.

DP and CP~\cite{2023sequenceparallelism,2023deepspeedulysses,2023ringattention} partition the activations along the token dimension across $c$ GPUs (Figure~\ref{fig:gtp-full-sharding}, top).
Each GPU computes on its own tokens, and the parameter gradients are all-reduced once per optimizer step, which overlaps with the backward pass when gradients are reduced in buckets~\cite{2020pytorchddp}.
Token partitions that contain whole sequences give DP; partitions that split sequences give CP, which matches DP for linear layers but requires the GPUs that share a sequence to communicate in attention and state-space layers.

In GTP, these are two independent partitioning axes, so a layer is distributed over $n = t \cdot c$ GPUs; they compose cleanly because they act on different inputs~\cite{2021megatronscaling}.
Partitioning alone leaves some state replicated: each TP partition is replicated across the $c$ GPUs that share it, and each GPU in a $t$-group holds the same \emph{boundary activations}, the activations at the input and output of each tensor-parallel region (such as the MLP block's input $X$).
Either replication can be removed by further sharding the replicated tensor and \emph{re-materializing} it with an all-gather just before it is used (Figure~\ref{fig:gtp-full-sharding}, bottom).
Re-materializing boundary activations across the $t$-group is SP~\cite{2023activationrecomputation}, which replaces each region's all-reduce with an all-gather before it and a reduce-scatter after it.
Re-materializing parameters across the $c$-group is the parameter sharding of FSDP and ZeRO-3~\cite{2020zero,2023pytorchfsdp}: parameters are all-gathered from their shards before use and freed afterward, in both the forward and backward passes of every microbatch, and gradients are reduce-scattered back to the shards.
Unlike activation all-gathers, these parameter all-gathers do not depend on preceding computation, so they can be issued ahead of the GEMMs that consume them.

Re-materialization need not span the whole partitioning group.
Parameters can be sharded and re-materialized within a \emph{weight re-materialization group} (\emph{weight group} for short) of $r$ of the $c$ GPUs: each GPU persistently stores $1/(t \cdot r)$ of each weight, and the weight group all-gathers its full TP partition ($1/t$ of the weight) just before each GEMM that consumes it.
The $c/r$ weight groups hold identical replicas, which synchronize gradients once per optimizer step, as in DP.
At $t=1$, setting $r = c$ gives FSDP, $r = 1$ gives DP, and intermediate values give hybrid sharding (HSDP)~\cite{2023pytorchfsdp}; larger $t$ composes these choices with TP.
The per-microbatch all-gathers then stay within the weight group, which can be placed on a fast interconnect such as NVLink~\cite{2026gb300nvl72}, while the once-per-step replica synchronization crosses slower links (Section~\ref{subsec:gtp-weight-data-views}).

Each weight is thus sharded across $p$ GPUs, which we call the \emph{GTP degree}: \begin{equation}
    \label{eq:gtp-degree}
    p = \underbrace{t}_{\text{TP degree}}\times
        \underbrace{r}_{\text{re-materialization degree}}.
\end{equation}
For example, GTP4 with TP2 ($t=2$, $r=2$) splits each weight into two TP partitions and stores each partition as two shards, which are all-gathered within the weight group before use.
With $r=1$, GTP reduces to TP with DP; with $t=1$, GTP reduces to FSDP or HSDP.
Table~\ref{tab:gtp-parallelism-schemes} places the familiar schemes in this space.

\begin{table}[!htbp]
\centering
\small
\begin{tabular}{@{}lcccc@{}}
\toprule
\multirow{2}{*}{\textbf{Scheme}} & \multirow{2}{*}{\textbf{\boldmath$t$}} & \multirow{2}{*}{\textbf{\boldmath$c$}} & \textbf{Re-materialize} & \textbf{Re-materialize} \\
& & & \textbf{activations} & \textbf{parameters} \\
\toprule
DP / CP & $1$ & $n$ & -- & -- \\
TP & $n$ & $1$ & -- & -- \\
TP+SP & $n$ & $1$ & \checkmark & -- \\
FSDP / ZeRO-3 & $1$ & $n$ & -- & \checkmark{} ($r = c$) \\
HSDP & $1$ & $n$ & -- & \checkmark{} ($1 < r < c$) \\
GTP & $t$ & $c$ & \checkmark & \checkmark{} ($r \le c$) \\
\bottomrule
\end{tabular}
\caption{Parallelism schemes as choices of partitioning and re-materialization.}
\label{tab:gtp-parallelism-schemes}
\end{table}

\paragraph{Mixture-of-experts layers.}
\phantomsection
\label{sec:moe}
MoE models~\cite{2017moe,2022switchtransformer,2024deepseekv3,2025llama4} increase model capacity without activating every parameter for each token: an MoE block replaces the MLP with $E$ expert MLPs and a router that sends each token to $k$ of them.
Expert parallelism (EP)~\cite{2021gshard,2026mcoremoetraining} partitions whole experts across GPUs and routes each token to the GPUs holding its experts and back with a pair of all-to-all collectives.
In an MoE block, EP's dispatch and combine all-to-alls take the place of TP+SP's activation all-gather and reduce-scatter, but exchange only the tokens selected by the routing mask.
Expert GTP (EGTP) applies GTP's parameter sharding and re-materialization to routed expert weights.
With EGTP degree $r_e$, each GPU in a weight group for expert weights stores $1/r_e$ of those weights and all-gathers the full weights before each GEMM that consumes them.
Dense GTP and EGTP degrees are configured independently.
Dense linear layers, including attention projections, use the composition of TP, SP, and GTP described above, while small parameters such as router weights remain replicated.

\begin{figure}[!htbp]
    \centering
    \includegraphics[width=\linewidth]{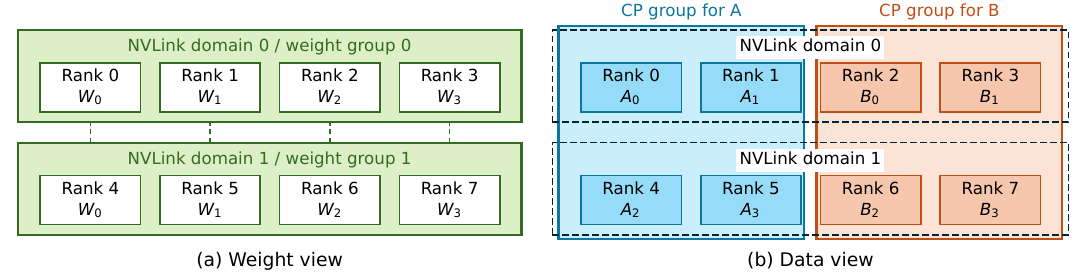}
    \caption{Independent weight and data views on eight GPUs (TP=PP=1, microbatch size 1).
    (a) Each NVLink domain stores a 4-way-sharded replica of $W$; weight all-gather (AG) and gradient reduce-scatter (RS) stay within the domain.
    Dashed links synchronize replicas of the same shard.
    (b) Samples $A$ and $B$ each span four sequence partitions in CP groups $\{0,1,4,5\}$ and $\{2,3,6,7\}$, respectively; attention communication crosses domains. %
    This small layout is only meant for illustrative purposes: in practice, CP groups are best kept within an NVLink domain when possible.
    }
    \label{fig:gtp-weight-data-views}
\end{figure}

\paragraph{Long sequences.}
\phantomsection
\label{par:gtp-long-sequences}
GTP can split a single long sequence across many GPUs without TP.
Linear layers process each token independently, so the GPUs in a weight group can each work on a different part of the same sequence.
A sequence can also span several weight groups (Figure~\ref{fig:gtp-weight-data-views}).
Splitting a sequence across GPUs does not change which GPUs store each weight shard (Section~\ref{subsubsec:gtp-batch-sequence}).
Weight all-gathers stay within each weight group, while attention and state-space layers communicate across the sequence's CP group.

\subsection{Communication}
\label{subsec:gtp-communication-analysis}

The choice of $t$ and $c$ trades parameter communication, which shrinks as $t$ grows, against activation communication, which grows with $t$.

To quantify this tradeoff, we model the MLP block's communication volume in tensor elements per GPU, for one microbatch of $S$ tokens per optimizer step and, when parameters are re-materialized, full parameter sharding ($r=c$).
The coefficients $\gamma_W$ and $\gamma_X$ scale the parameter and activation volumes relative to TP+CP without re-materialization, for which $\gamma_W = \gamma_X = 1$.
Without re-materialization, the MLP block uses a gradient all-reduce over the $c$-group and two activation all-reduces over the $t$-group, one in the forward pass and one in the backward pass.
Parameter re-materialization replaces the gradient all-reduce with two parameter all-gathers (one for each pass) and a gradient reduce-scatter, giving $\gamma_W = 1.5$.
Activation re-materialization uses SP's all-gather and reduce-scatter in each pass, plus an additional all-gather of the inputs to the weight-gradient GEMM, giving $\gamma_X = 1.25$: \begin{equation}
    \label{eq:gtp-communication}
    \text{comm}(t, c) = \gamma_W\,\frac{4hh'}{t}\left(1-\frac{1}{c}\right)
        + \gamma_X\,\frac{4Sh}{c}\left(1-\frac{1}{t}\right).
\end{equation}
With $r<c$, the per-microbatch parameter all-gathers and gradient reduce-scatters stay within the weight group of $r$ GPUs, and the $c/r$ replicas additionally all-reduce gradients once per optimizer step.
If collectives use different precisions, each must be weighted by its bytes per element.
Under gradient accumulation, parameter all-gathers repeat for each microbatch unless materialized parameters are retained across microbatches at a memory cost~\cite{2026fsdp2api}.

For the single-microbatch model in Equation~\ref{eq:gtp-communication}, let $s = S/n$ be each GPU's share of the microbatch's tokens.
With large $t$- and $c$-groups, the parameter term is approximately $4\gamma_W hh'/t$ and the activation term approximately $4\gamma_X sht$.
Dividing these terms by their respective group bandwidths, $B_c$ and $B_t$, gives a bandwidth-only communication-time model when all collectives use the same precision.
For $\gamma_W=1.5$, $\gamma_X=1.25$, and fixed $s$, the modeled communication time is minimized at \begin{equation}
    \label{eq:gtp-tp-balance}
    t^* = \sqrt{1.2 \cdot (h'/s) \cdot (B_t/B_c)}.
\end{equation}
Many tokens per GPU push $t^*$ toward $1$ (FSDP); few tokens per GPU favor larger $t$ within an NVLink domain.

At fixed $s$, TP's activation volume per GPU is proportional to $t-1$ (Equation~\ref{eq:gtp-communication}), since each GPU's GEMMs consume the activations of all $s \cdot t$ tokens in its $t$-group.
In contrast, for MoE models, each token is sent to at most $k$ GPUs (those hosting the experts the token is routed to), so a GPU's EP volume is at most $k$ times its tokens regardless of the EP group size.
EP groups can thus be much larger than TP groups.

The communication-time model above ignores two effects: any overlap of communication with computation, and the latency term of each collective (small for the message sizes typical in training).
Overlap matters most for parameter all-gathers, which can be issued ahead of the GEMMs that consume them (Section~\ref{subsec:gtp-framework}).
We next estimate how much computation is available to hide them.

\paragraph{Hiding all-gathers behind computation.}
More tokens per expert add computation without increasing weight all-gather volume.
Consider one expert GEMM at TP1.
Balanced routing across an expert-parallel group of $\mathrm{EP}$ GPUs gives each expert $s_e = \mathrm{EP}\cdot s\cdot k/E$ tokens.
Its weight has $hh'$ elements of $b$ bytes each, sharded across $r_e>1$ GPUs, where $h'$ is the expert's intermediate size.
At achieved GEMM throughput $F$ (FLOP/s) and all-gather bandwidth $B_r^e$ (bytes/s), the ratio of one GEMM's time to one weight all-gather's time is
\begin{equation}
    \label{eq:gtp-overlap-ratio}
    \frac{t_{\mathrm{GEMM}}}{t_{\mathrm{AG}}}
    = \frac{2s_e hh'/F}{b(1-1/r_e)hh'/B_r^e}
    = \frac{2s_e B_r^e}{b(1-1/r_e)F}.
\end{equation}
For fixed $F$ and $B_r^e$, the weight size $hh'$ cancels, so the ratio depends on tokens per expert rather than weight size; in practice, $F$ also depends on GEMM shape.
For dense layers at TP1, each GPU processes $s$ tokens and all-gathers weights across its $r$-GPU weight group; the same analysis applies.
A higher ratio offers more opportunity to hide communication; actual overlap also depends on collective startup costs, network contention, and scheduling (Section~\ref{subsec:gtp-scheduling}).

\subsection{Memory}
\label{subsec:gtp-memory}

Partitioning parameters (TP) shrinks per-GPU parameter memory, and partitioning tokens (DP or CP) shrinks per-GPU activation memory; re-materialization extends both reductions to all $n$ GPUs.
For the MLP block without re-materialization, per-GPU parameter storage is $2hh'/t$ and boundary-activation storage is $Sh/c$, counted in tensor elements and excluding intermediate activations.
Re-materializing parameters over $r$ GPUs reduces the former to $2hh'/(tr)$, or $2hh'/n$ when $r = c$; re-materializing activations reduces the latter to $Sh/n$.
Gradient sharding follows parameter sharding, and the distributed optimizer partitions optimizer state across all $n$ GPUs.
Peak memory also includes re-materialized parameters and communication buffers, whose sizes depend on how far ahead all-gathers are issued (Section~\ref{sec:design}).

%% file: sections/3_parameter_scheduling.tex
\section{Per-GEMM Scheduling of Parameter Re-materialization}
\label{sec:design}

Section~\ref{sec:gtp} treats parameter re-materialization as a preprocessing step on one input of a GEMM.
Building on this view, this section describes how we schedule re-materialization per GEMM, using a consumer interface and policies that coordinate each all-gather with the GEMM that consumes it.
A weight's \emph{consumers} are the GEMMs that read it, and its \emph{view} is the re-materialized copy that they read.
Each expert is an MLP with two linear layers, FC1 and FC2; with grouped GEMMs, a GPU computes FC1 for all of its local experts in one GEMM (GFC1) and FC2 in another (GFC2).

\subsection{Scheduling Problem}
\label{subsec:gtp-fsdp-problems}

As discussed in Section~\ref{subsec:gtp-framework}, parameter all-gathers, unlike activation all-gathers, can be issued ahead of the GEMMs that consume them.
Any such schedule must balance three coupled concerns, especially in MoE models.

\paragraph{P1: Compute-communication overlap.}
Each all-gather must be issued early enough to finish before its consumer, and the consumer should not wait for unrelated communication.
Consumers inside one module can need their parameters at different times: in a fused MoE layer, GFC1 and GFC2 are separate GEMMs, so if both wait for the same set of all-gathers, the GFC1 GEMM is delayed by the all-gather of the GFC2 weights.
Expert parameters sharded across NVLink domains need longer prefetch windows because their all-gathers use slower inter-domain InfiniBand~\cite{2026infiniband} or RoCE~\cite{2026roce} links.

\paragraph{P2: Memory footprint.}
Re-materialization shrinks each GPU's persistent weight memory to one stored shard ($1/(t \cdot r)$ of each weight).
However, each view holds the full TP partition ($1/t$ of the weight) regardless of $r$.
Each full gradient also waits in a buffer until its reduce-scatter has read it.
Peak memory therefore depends on how many views and gradient buffers the schedule keeps live: prefetching further ahead adds views, and overlapping gradient reduction holds buffers longer.
A schedule should materialize each parameter as late as possible and release it as soon as its last reader finishes.

\paragraph{P3: Collective efficiency.}
All-gathering parameter tensors separately allows each all-gather to be scheduled independently, but each collective incurs launch overhead and communication startup latency.
For small tensors, these fixed costs can dominate data transfer time, making the all-gathers latency-bound.
\emph{Coalescing} combines the all-gathers of several tensors into one collective, paying launch overhead once and amortizing startup costs over a larger transfer.
However, combining parameters needed by different consumers can delay early consumers and keep more views live.
Grouped GEMMs illustrate this tradeoff: the local expert weights of one grouped GEMM are consumed together and can share a collective, whereas GFC1 and GFC2 need their weights at different times.

\begin{figure}[!htbp]
    \centering
    \includegraphics[width=\linewidth]{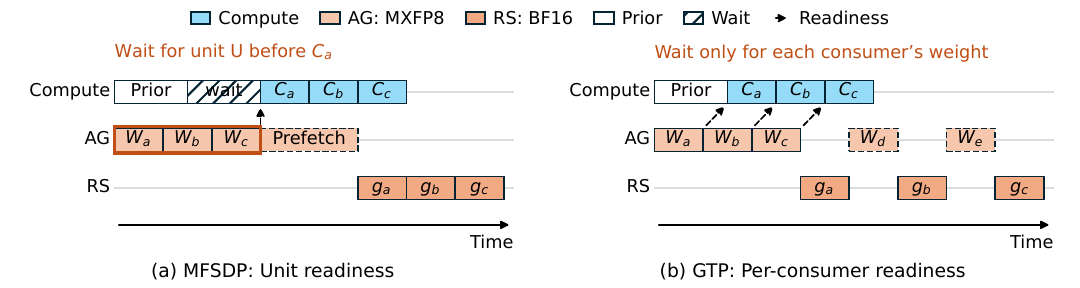}
    \caption{Backward-pass scheduling (schematic): (a) MFSDP waits per unit; (b) GTP waits per consumer.
    $C_i$, $W_i$, and $g_i$ denote a consumer, its weight, and its gradient.
    In both panels, AG and RS share one communicator over the same GPUs and execute sequentially.
    }
    \label{fig:gtp-mfsdp-backward-schedule}
\end{figure}

\paragraph{Scheduling in existing FSDP implementations.}
Existing FSDP implementations schedule re-materialization through module hooks, typically on Transformer blocks (Section~\ref{sec:related}).
A hook waits for all of its module's parameter all-gathers before the module runs.
With block-level hooks, a block's first GEMM therefore waits for every parameter all-gather in the block (P1; Figure~\ref{fig:gtp-mfsdp-backward-schedule}), and temporary memory and collective grouping follow the block (P2--P3).
Hooks on smaller modules are possible, but a module hook cannot reach inside a fused operator: when expert weights are sharded, a fused expert layer's GFC1 GEMM also waits for GFC2's weights.
We instead schedule re-materialization separately for each consumer.
Table~\ref{tab:gtp-scheduling-boundaries} compares how PyTorch FSDP2, MFSDP, and GTP schedule parameter all-gathers.

\begin{table}[!htbp]
    \centering
    \begingroup
    \small
    \setlength{\tabcolsep}{6pt}
    \renewcommand{\arraystretch}{1.12}
    \begin{tabularx}{\linewidth}{@{}>{\raggedright\arraybackslash}p{0.13\linewidth}
        >{\raggedright\arraybackslash}X
        >{\raggedright\arraybackslash}X
        >{\raggedright\arraybackslash}X@{}}
        \toprule
        & \textbf{PyTorch FSDP2}~\cite{2026fsdp2schedsource,2026torchtitanschedsource} & \textbf{MFSDP}~\cite{2026mfsdpbaselinebase} & \textbf{GTP} \\
        \toprule
        \textbf{Scheduling granularity} & Module wrapped by \texttt{fully\_shard} (typically a Transformer block). &
        Transformer block or finer submodule (configurable). & Individual GEMM. \\
        \addlinespace
        \textbf{When to issue (\hyperref[subsubsec:gtp-dense-overlap]{C1})} & Next module, in execution order or by explicit target. &
        Next parameter groups in configured order, within a prefetch budget. & Per chain: next dense GEMM; next MoE block for experts. \\
        \addlinespace
        \textbf{What a GEMM waits for (\hyperref[subsubsec:gtp-precise-waits]{C2})} & All all-gathers of its module. &
        All all-gathers of the hooked module (a Transformer block or a smaller submodule). & Only its own weights. \\
        \addlinespace
        \textbf{Fused MoE layer} & GFC1 also waits for GFC2's weights when sharded. &
        GFC1 also waits for GFC2's weights when sharded. & GFC1 and GFC2 each wait for their own weights. \\
        \bottomrule
    \end{tabularx}
    \endgroup
    \caption{How each implementation schedules parameter all-gathers.
    Section~\ref{subsec:gtp-scheduling-tradeoffs} defines C1, C2, and GTP's prefetch chains.
    Hook granularity is configurable in both FSDP implementations; Section~\ref{subsubsec:gtp-eval-mfsdp-scaling} describes the evaluated MFSDP configurations.}
    \label{tab:gtp-scheduling-boundaries}
\end{table}

\subsection{Consumer Interface}
\label{subsec:gtp-weight-layout}
\label{subsec:gtp-weight-abstraction}
\label{par:gtp-weight-abstraction}

\begin{figure}[!htbp]
    \centering
    \includegraphics[width=\linewidth]{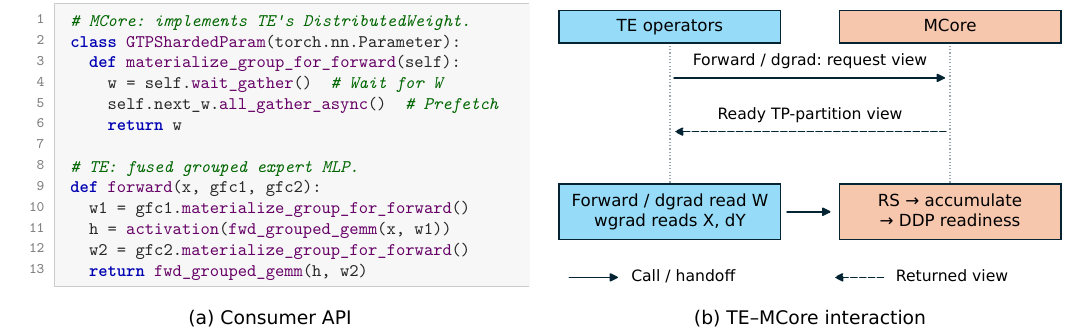}
    \caption{Consumer interface between Megatron-Core (MCore) and Transformer Engine (TE).
    (a) TE requests weight views for GFC1 and GFC2; MCore waits for each
    all-gather and prefetches a later consumer's weight.
    (b) The forward and dgrad GEMMs consume weight views; wgrad uses $X$ and $dY$ and hands
    gradients to MCore for reduce-scatter and accumulation.
    Startup and buffer management are omitted.}
    \label{fig:gtp-api}
    \label{fig:gtp-call-stack}
\end{figure}

Each GEMM that reads a sharded weight obtains its view through a common consumer interface rather than through module hooks (Figure~\ref{fig:gtp-api}).
The forward, activation-recomputation, and input-gradient (dgrad) GEMMs read $W$.
The weight-gradient (wgrad) GEMM reads only the layer's input activations and output gradients, so the view can be released before wgrad.
The consumer then hands the weight gradient back to be reduce-scattered into the stored shards.

A GEMM waits for its weight's all-gather to complete, then issues the all-gather for a later GEMM's weight, and then starts its computation.
The scheduling policies in Section~\ref{subsec:gtp-scheduling-tradeoffs} determine the prefetch target and collective grouping.
Section~\ref{sec:implementation} describes how TE and MCore implement this interface.

\subsection{Scheduling Policies}
\label{subsec:gtp-scheduling-tradeoffs}
\label{subsec:gtp-scheduling}
\label{subsec:gtp-scheduling-details}

Scheduling all-gathers per GEMM involves three decisions: when to issue each all-gather (C1), what each GEMM waits for (C2), and which parameters share a collective (C3).
These decisions matter most in MoE layers, where GFC1 and GFC2 run inside one fused module and expert all-gathers can cross NVLink domains.

\begin{figure}[!htbp]
    \centering
    \includegraphics[width=\linewidth]{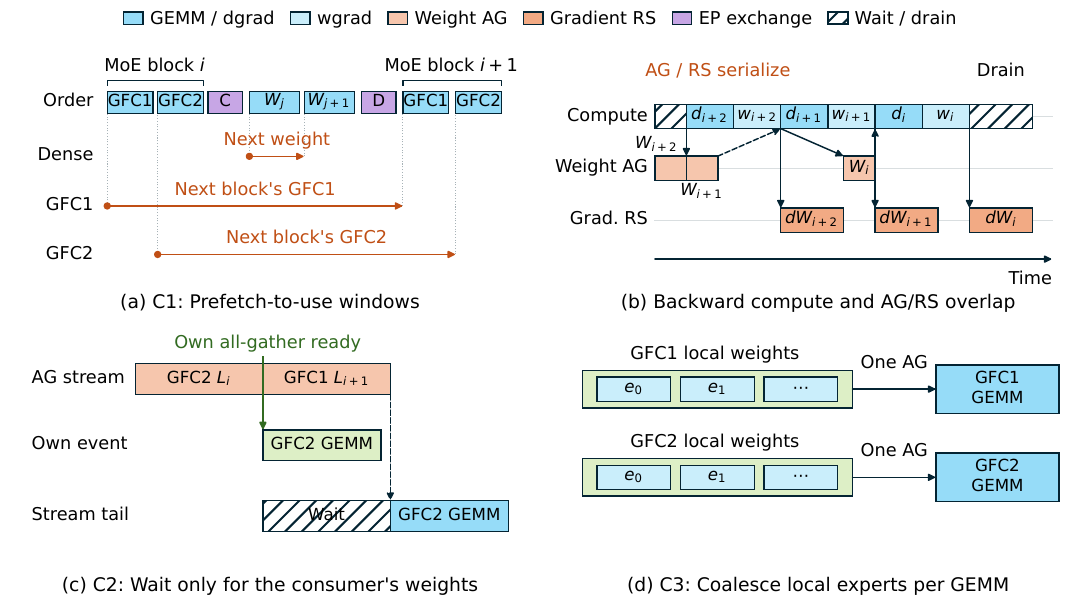}
    \caption{GTP's three parameter-scheduling policies.
    (a) C1: next-weight dense and next-block expert prefetching.
    (b) Backward-pass overlap with serialized AG and RS.
    (c) C2: per-consumer waits.
    (d) C3: per-GEMM expert coalescing.
    D and C denote EP dispatch and combine; $d$ and $w$ denote dgrad and wgrad; $L_i$ and $e_j$ denote a layer and an expert weight.}
    \label{fig:gtp-scheduling-policies}
    \label{fig:gtp-dense-overlap}
    \label{fig:gtp-prefetch-window}
\end{figure}

\paragraph{When to issue collectives (C1).}
\phantomsection
\label{subsubsec:gtp-dense-overlap}
\label{subsubsec:gtp-projection-chains}
\label{para:gtp-expert-overlap}
We choose the lookahead to balance communication overlap against the memory occupied by prefetched weights.
A \emph{prefetch chain} specifies which weight to all-gather next when a GEMM consumes the current weight.
Chain membership determines the lookahead; Section~\ref{subsec:gtp-chain-initialization} describes registration and warmup.
\begin{itemize}[leftmargin=*, itemsep=4pt, topsep=4pt]
\item \textbf{Dense weights: hide fast all-gathers with short lookahead.}
When the weight group for dense weights lies within an NVLink domain, its all-gathers are fast, so a short lookahead suffices and limits the memory held by prefetched views.
By default, all dense weights share one chain: consuming $W_i$ prefetches $W_{i+1}$ in the forward pass and $W_{i-1}$ in the backward pass (Figure~\ref{fig:gtp-scheduling-policies}(a,b)).
Whether the all-gather of $W_{i+1}$ is fully hidden depends on its duration relative to the compute time of the GEMM on $W_i$.

\item \textbf{Expert weights: hide slower all-gathers with longer lookahead.}
Cross-domain EGTP all-gathers need a longer overlap window (Section~\ref{subsec:gtp-cross-domain-experts}), so GFC1 and GFC2 use separate default chains.
GFC1 and GFC2 each prefetch the corresponding weights in the next MoE block, allowing their all-gathers to overlap the rest of the current block, intervening dense layers, and EP token exchange (Figure~\ref{fig:gtp-scheduling-policies}(a)).
This longer window provides more computation to hide the slower all-gathers (P1), but keeps prefetched views live longer (P2).
\end{itemize}

In the backward pass, each layer runs dgrad before wgrad, and wgrad does not read the weight.
Issuing the next all-gather before dgrad thus lets it overlap both GEMMs (Figure~\ref{fig:gtp-scheduling-policies}(b)).
Once wgrad finishes, its gradient is reduce-scattered, which can overlap the GEMMs of the layers that the backward pass processes next.
Each GPU keeps its reduced gradient shards until the optimizer step.

\paragraph{What each GEMM waits for (C2).}
\phantomsection
\label{subsubsec:gtp-precise-waits}
\label{para:gtp-completion-waits}
Our implementation records a completion event after each all-gather, before any later prefetch is queued, and each GEMM waits only on its own event (Figure~\ref{fig:gtp-scheduling-policies}(c)).
FSDP2 also records an event per all-gather~\cite{2026fsdp2schedsource}, but its module hooks wait for every all-gather in the configured unit.
When GFC1 and GFC2 share a unit, the GFC1 GEMM also waits for GFC2's weights.

\paragraph{Which parameters share a collective (C3).}
\phantomsection
\label{subsubsec:gtp-expert-coalescing}
The local expert weights feeding one grouped GEMM are consumed together, so they are all-gathered in a single coalesced collective, with separate groups for GFC1 and GFC2 (Figure~\ref{fig:gtp-scheduling-policies}(d)).
This grouping improves collective efficiency (P3).
Per-expert all-gathers require a separate collective for each local expert, while grouping GFC1 and GFC2 together would couple consumers that need their weights at different times.
Section~\ref{subsec:gtp-eval-ablations} evaluates the throughput benefit of this coalescing policy.

%% file: sections/4_implementation.tex
\section{Implementation}
\label{sec:implementation}
\label{subsec:gtp-implementation-details}

\begin{figure}[!htbp]
    \centering
    \includegraphics[width=0.90\linewidth]{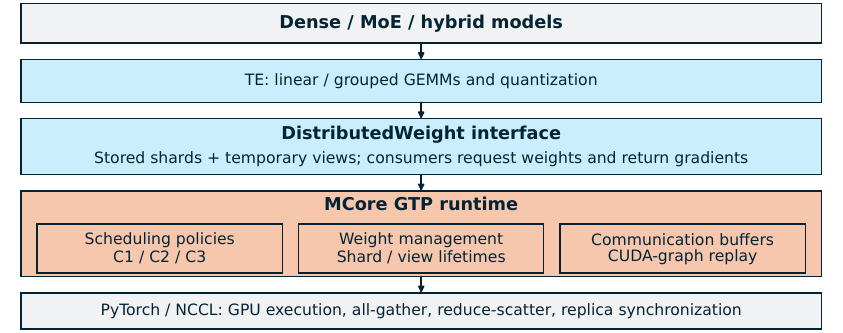}
    \caption{GTP architecture. TE consumes views through \texttt{DistributedWeight};
    MCore manages scheduling (\S\ref{subsec:gtp-scheduling-tradeoffs}),
    weights (\S\ref{subsec:gtp-weight-storage}), buffers (\S\ref{subsec:gtp-symmetric-buffers}),
    and graph replay (\S\ref{subsec:gtp-cuda-graphs}).
    The runtime builds on PyTorch and uses NCCL for collective communication.
    Colors distinguish models and infrastructure (gray), computation and interfaces (blue), and the GTP runtime (orange).}
    \label{fig:gtp-system-overview}
\end{figure}

This section describes how we implement GTP and the scheduling policies of Section~\ref{sec:design} in Megatron-Core (MCore)~\cite{2026gtpimplementation} and Transformer Engine (TE)~\cite{2026transformerengine}.
Figure~\ref{fig:gtp-system-overview} summarizes the implementation: TE operators consume weight views, while MCore manages sharded storage, communication, prefetch scheduling, and CUDA graph replay.
TE defines the consumer interface of Section~\ref{subsec:gtp-weight-abstraction} as the \texttt{DistributedWeight} protocol, which MCore's \texttt{GTPShardedParam} implements.
TE operators obtain weights and return gradients through functions that accept both distributed weights and ordinary tensors, so the same operator code runs with or without GTP.
When MCore constructs a TE linear or grouped-linear operator, it passes the padded row count of the stored shard, so the operator never allocates the full weight.

\FloatBarrier

\subsection{GPU Groups and Placement}
\label{subsec:gtp-composition}
\label{subsec:gtp-weight-data-views}

MCore organizes GPUs into groups, each a set of GPUs that run collectives together.
Each GPU belongs to a TP group (GPUs that hold different TP partitions of each weight), a weight group (GPUs that each store a different shard of the same TP partition), a replica group (GPUs that hold the same shard), and a CP group (GPUs that share a sequence).

\subsubsection{Weight Groups}
\label{subsubsec:gtp-weight-groups}

GTP splits each TP partition ($1/t$ of a weight) into $r$ shards ($1/(t \cdot r)$ of the weight), one per GPU in the weight group.
Each GPU stores only its shard and all-gathers the full TP partition just before each GEMM that uses it.
GTP splits each TP partition along its rows (the output dimension, in the \texttt{[out, in]} layout that PyTorch and TE use for weights).
In a column-parallel layer, TP and GTP thus split the same dimension; in a row-parallel layer, TP splits the input dimension and GTP the output dimension.
Weight gradients are reduce-scattered within the weight group.
MCore's \texttt{DistributedDataParallel} wrapper (DDP) then synchronizes gradients across replicas of each shard, and MCore's distributed optimizer partitions optimizer state across these replicas.

\label{par:gtp-selective-sharding}
Sharding does not pay off for every weight.
A small weight saves little memory when sharded, but it still needs an all-gather before each GEMM that uses it.
It can also need large padding: DSV4-Pro's KV projection has only 512 rows, so sharding it across 64 GPUs under MXFP8~\cite{2023microscaling} would pad it to four times its size (Sections~\ref{subsec:gtp-padding} and~\ref{subsec:gtp-eval-padding}).
GTP can instead keep such weights replicated: each GPU then stores the full TP partition, and the weight does not need an all-gather or padding.
For example, our DSV4-Pro runs keep the KV projection replicated.

\subsubsection{Splitting Sequences Within and Across Weight Groups}
\label{subsubsec:gtp-batch-sequence}
\label{subsubsec:gtp-cp-folding}

A long sequence may need to be split across more GPUs than one weight group contains.
Instead of enlarging the weight group, which would push weight all-gathers across NVLink domains, GTP splits the sequence across several weight groups.
In a $q\times m$ placement, each sequence is split across $q$ GPUs in each of $m$ weight groups, for an effective CP degree of $qm$.
We call $m$ the outer CP degree.
Within each weight group of size $r$, the remaining factor $r/q$ splits independent sequences along the batch dimension.
Figure~\ref{fig:gtp-weight-data-views}(b) illustrates the $2\times2$ case: each sequence spans two GPUs in each of two weight groups.

For example, the 1M-token configuration in Table~\ref{tab:internal2t-long-context-configs} uses $r=q=64$ and $m=2$.
Each sequence spans two 64-GPU weight groups, one per NVLink domain, giving CP128 with GTP64 and no batch split within a weight group ($r/q=1$).
Weight all-gathers use NVLink, while CP communication crosses domains.
Each GPU stores $1/64$ of each sharded dense weight, twice the $1/128$ it would store with a single 128-GPU weight group.

\subsubsection{Expert Weight Sharding Across NVLink Domains}
\label{subsec:gtp-cross-domain-experts}

We place each EP group within one NVLink domain (64 of a domain's 72 GPUs~\cite{2026gb300nvl72}), so EP's dispatch and combine communication stays on NVLink.
Without EGTP, the GPUs at the same position in every EP group hold identical expert weights.
EGTP shards these weights across $r_e$ such GPUs in different domains, reducing each GPU's expert-weight storage roughly $r_e$-fold.
The expert all-gathers then cross the slower inter-domain network.
The separate GFC1 and GFC2 chains give each of these all-gathers a full MoE block to complete (Section~\ref{subsubsec:gtp-projection-chains}).

\subsection{Weight Storage and Communication Buffers}
\label{subsec:gtp-weight-storage}

Each GPU stores its weight shards persistently and needs temporary buffers for the TP partitions it all-gathers and the gradients it reduce-scatters.
The format and size of these buffers determine both memory use and collective cost.

\subsubsection{Low-Precision Weights}
\label{subsec:gtp-low-precision}

GTP can store weight shards in MXFP8~\cite{2023microscaling} or NVFP4~\cite{2025nvfp4}, using TE's training recipes for these formats.
The optimizer quantizes each updated FP32 master shard once per step.
All-gathers then transfer the quantized shards and their scales directly, so no microbatch re-quantizes a weight.
Each stored shard holds two quantized layouts: row-wise for the forward-pass GEMM and column-wise for the dgrad GEMM.
Each all-gather transfers only the layout that its GEMM reads.
Low-precision storage thus reduces both persistent weight memory and all-gather traffic.
It does not reduce the precision of gradient communication (BF16 or FP32, configured separately), or of the master weights or optimizer state (both in FP32 by default).

\subsubsection{Alignment Padding}
\label{subsec:gtp-padding}

\begin{figure}[!htbp]
    \centering
    \begin{minipage}[b]{0.60\linewidth}
        \centering
        \includegraphics[trim=5bp 68.52bp 25bp 0bp,clip,width=\linewidth]{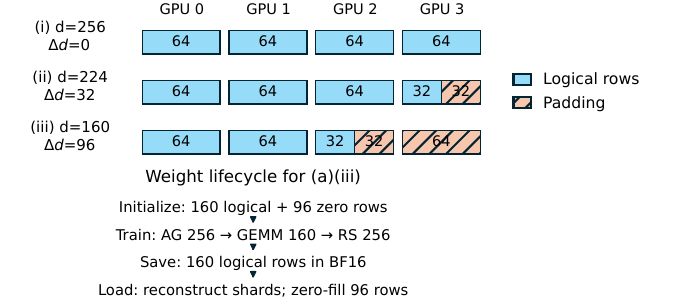}
        \par\small (a) Aligned shards
    \end{minipage}\hspace{0.01\linewidth}%
    \begin{minipage}[b]{0.34\linewidth}
        \centering
        \includegraphics[trim=35bp 0bp 128.6bp 79.07bp,clip,width=\linewidth]{figures/gtp_padding.pdf}
        \par\small (b) Weight lifecycle
    \end{minipage}
    \caption{(a) MXFP8 alignment of four stored shards (64 rows each).
    Cases (i--iii) show increasing padding; GPU 3 is fully padded in (iii).
    (b) Weight lifecycle for (a)(iii),
    excluding padding from GEMMs and checkpoints.}
    \label{fig:gtp-padding}
\end{figure}

Each GPU stores one shard of its TP partition of each weight.
For a TP partition ($1/t$ of a weight) with $d$ rows split across a weight group of $r$ GPUs, each stored shard ($1/(t \cdot r)$ of the weight) nominally holds $d/r$ rows.
MXFP8 and NVFP4 require each stored shard's row count to be a multiple of an alignment $a$, which is $32$ for MXFP8 and $16$ for NVFP4 ($a=1$ for BF16).
Each GPU quantizes only its own stored shard (Section~\ref{subsec:gtp-low-precision}), so no quantization block may span two stored shards.
These alignments keep shard boundaries on quantization-block boundaries, allowing GTP to all-gather quantized weights and their scales directly without dequantizing and re-quantizing them.
When $d/r$ is not such a multiple, GTP appends zero rows to the TP partition, so that every stored shard has the same aligned row count.
The padded TP partition has $\widetilde d$ rows, of which $\Delta d$ are padding, and each stored shard has $d_{\mathrm{local}}$ rows:
\begin{equation}
    \label{eq:gtp-alignment-padding}
    \widetilde d = ar\left\lceil\frac{d}{ar}\right\rceil,\qquad
    \Delta d = \widetilde d-d,\qquad
    d_{\mathrm{local}} = \frac{\widetilde d}{r}.
\end{equation}
An all-gather thus reconstructs a TP partition of $\widetilde d$ rows: the $d$ logical rows followed by $\Delta d$ padding rows (Figure~\ref{fig:gtp-padding}(a)).

GEMMs use only the logical rows.
Before the forward-pass, dgrad, or recomputation GEMM, GTP trims the padding from the all-gathered TP partition and adjusts its scales without dequantizing it.
The wgrad GEMM computes gradients for only the logical rows, and GTP places them in a zero-padded buffer of $\widetilde d$ rows, which is reduce-scattered back into stored shards.
Checkpoints also store only the logical rows (Figure~\ref{fig:gtp-padding}(b)).
Loading a checkpoint recomputes the padding for the target sharding configuration (Section~\ref{subsec:gtp-checkpointing}).
Padding thus adds storage and communication but no computation.
Section~\ref{subsec:gtp-eval-padding} quantifies this overhead across benchmark models.

FSDP2 also pads its shards when a weight does not divide evenly~\cite{2026fsdp2impl}, but not to the alignment that low-precision formats require.
In TE's FSDP2 integration, unsupported MXFP8 splits fall back to high precision~\cite{2026tefsdp2}.
NVFP4 all-gathers there reject shards whose row count is not divisible by 16.

\subsubsection{Symmetric User Buffers}
\label{subsec:gtp-symmetric-buffers}
\begin{figure}[!htbp]
    \centering
    \includegraphics[width=\linewidth]{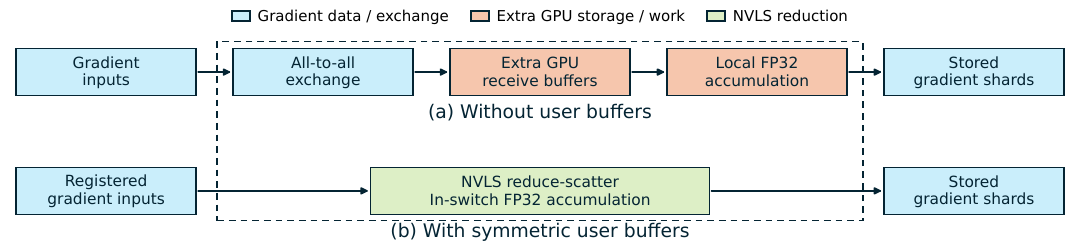}
    \caption{Gradient reduction with and without symmetric user buffers.
    Both paths communicate BF16 gradients and accumulate in FP32.
    The dashed region shows how NVLS replaces the all-to-all exchange, extra receive buffers, and local FP32 accumulation with a single reduce-scatter.}
    \label{fig:gtp-user-buffer-paths}
\end{figure}

As discussed in Section~\ref{subsec:gtp-communication-analysis}, re-materialization trades off extra communication for reduced memory footprint.
One way of reducing communication cost is to use a one-shot algorithm for the reduce-scatter (an all-to-all exchange followed by local accumulation), which allows reduced communication precision over the wire with FP32 accumulation (Figure~\ref{fig:gtp-user-buffer-paths}(a)).
However, this requires additional receive buffers (one per peer GPU), memory traffic, and streaming-multiprocessor (SM) work.
Registering buffers with NCCL as symmetric user buffers (UB) instead enables NVLink SHARP (NVLS) reduce-scatter with in-switch FP32 accumulation, avoiding those receive buffers and the separate local reduction~\cite{2026ncclbufferregistration,hamidouche2025gpu} (Figure~\ref{fig:gtp-user-buffer-paths}(b)).

GTP can register only the gradient reduce-scatter inputs (UB RS), or also the stored weight shards and all-gathered TP partitions (UB AG+RS).
Registering the weight buffers lets NCCL's all-gathers also access peer GPU memory directly, through NVLS or direct memory access (DMA)~\cite{hamidouche2025gpu}.
Section~\ref{subsec:gtp-eval-ub} compares the two modes' throughput and memory use.
Three mechanisms make symmetric user buffers practical:

\begin{itemize}[leftmargin=*, itemsep=4pt, topsep=4pt]
\item \textbf{Buffer registration.}
GTP allocates symmetric user buffers from one memory pool per weight group, with separate pools for dense and expert weights, and registers each pool with NCCL for its group.
Registration is collective: all GPUs in a group register their pools together.
GTP registers each pool during model construction, before the first forward pass.
When a pool grows, its new memory is registered automatically.

\item \textbf{Buffer reuse and synchronization.}
Each reduce-scatter input holds the full TP-partition gradient from a wgrad GEMM, $r$ times the size of the stored shard.
A dedicated registered input buffer for every weight would therefore use as much memory as unsharded gradients.
GTP reuses registered buffers through a last-in-first-out (LIFO) free list keyed by element count, dtype, and group.
Weights with the same element count can share a buffer even if their shapes differ.
On release, the buffer carries a completion event for its reduction.
The stream that next allocates the buffer waits on that event before overwriting it, without a device-wide synchronization.

\item \textbf{Direct gradient writes.}
Supported TE wgrad GEMMs write their output directly into the registered gradient-input buffer in Figure~\ref{fig:gtp-user-buffer-paths}(b), skipping an extra copy.
The buffer already includes the necessary padding rows.
Other GEMMs, and GEMMs whose output dtype differs from the buffer's, write to a temporary buffer that GTP then copies into the registered one.
\end{itemize}

\subsection{Prefetch-Chain Registration and Initialization}
\label{subsec:gtp-chain-initialization}

\begin{figure}[!t]
    \centering
    \includegraphics[width=\linewidth]{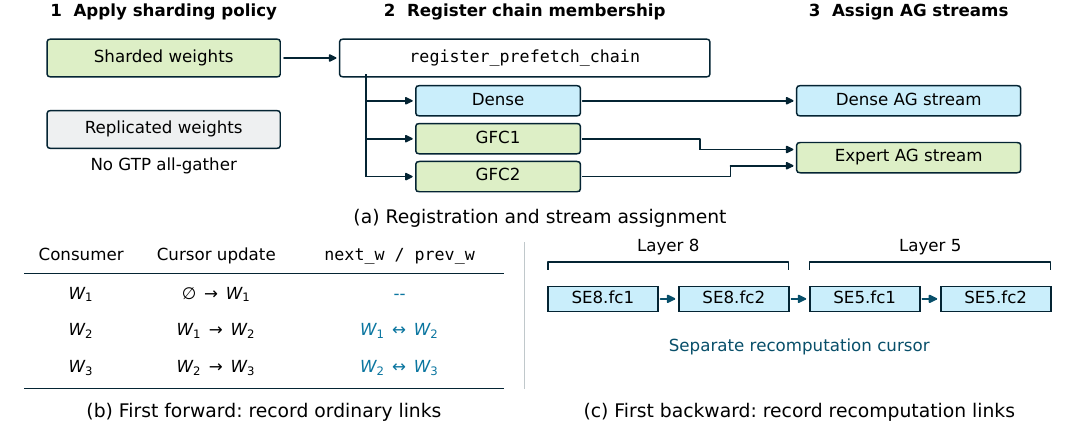}
    \caption{Prefetch-chain setup and warmup.
    (a) Sharded weights register with separate chains; GFC1 and GFC2 share an expert
    all-gather stream. (b) Forward-pass consumers record \texttt{next\_w}/\texttt{prev\_w}
    links and advance the cursor.
    (c) Shared-expert (SE) recomputation learns a separate order, illustrated
    by Nemotron-3 Super layers 8 then 5 (zero-based indices).}
    \label{fig:gtp-chain-registration}
    \label{fig:gtp-chain-init}
\end{figure}

Users assign weights to a chain by calling \texttt{register\_prefetch\_chain} with a chain name and a list of weights.
MCore then assigns every remaining sharded weight to the default chains described in Section~\ref{subsec:gtp-scheduling-tradeoffs}; replicated weights belong to no chain.
All weights in a chain must share one group, and a grouped-linear operator joins a chain with all of its local experts.
Chains on the same group can share a communication stream; GFC1 and GFC2 all-gathers, for example, share the expert stream (Figure~\ref{fig:gtp-chain-registration}(a)).

During the first forward pass, each chain records each weight's successor and predecessor (Figure~\ref{fig:gtp-chain-init}(b)).
The ordinary backward pass follows the predecessor links; activation recomputation needs a separate order.
It reruns forward-pass GEMMs during the backward pass, so it visits layers in reverse but the GEMMs within each layer in forward order, and only for the layers that recompute.
In Figure~\ref{fig:gtp-chain-init}(c), Nemotron-3 Super recomputes its shared experts in layer~8 and then layer~5, running FC1 before FC2 in each.
The first backward pass records this order separately, and recomputation all-gathers the row-wise layout into its own buffers, apart from the column-wise layout the dgrad GEMM uses.
CUDA graph replay requires every training step to visit each chain's weights in recorded order.

\subsection{Full-Iteration CUDA Graphs}
\label{subsec:gtp-cuda-graphs}

\begin{figure}[!htbp]
    \centering
    \includegraphics[width=\linewidth]{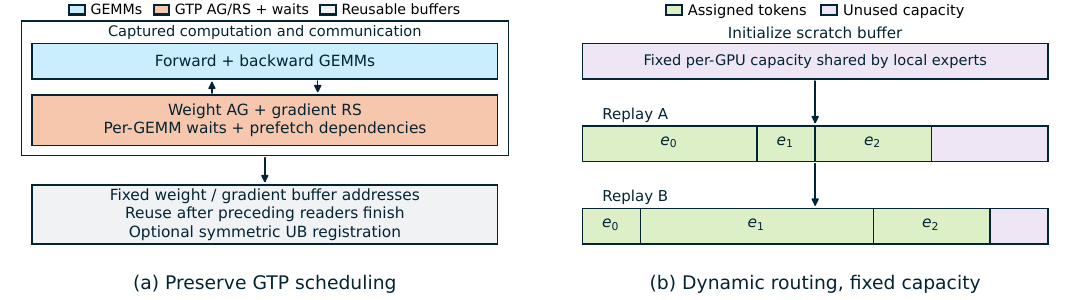}
    \caption{Full-iteration CUDA graphs with GTP.
    (a) Capture once; replay GTP dependencies with fixed buffer addresses.
    (b) Token counts vary across replays while buffer addresses and capacity stay fixed, avoiding host-side resizing; segment widths show token counts for local experts $e_i$.}
    \label{fig:gtp-full-iteration-graphs}
\end{figure}

GTP uses full-iteration CUDA graphs~\cite{2026cudagraphs} to reduce kernel-launch and host-scheduling overhead.
Capturing the forward and backward passes across a training step's microbatches requires preserving GTP's parameter schedule and accommodating MoE token counts that vary between training steps (Figure~\ref{fig:gtp-full-iteration-graphs}).

\paragraph{Preserving GTP scheduling during replay.}
\label{subsec:full-iter-cg}
Warmup establishes the prefetch order described in Section~\ref{subsec:gtp-chain-initialization} and provisions reusable weight and gradient buffers.
Capture records GEMMs, weight all-gathers, gradient reduce-scatters, and accumulation together with their stream and event dependencies.
Replay therefore preserves per-GEMM waits and communication overlap (Figure~\ref{fig:gtp-full-iteration-graphs}(a)).
Every buffer keeps its address across replays, and a buffer is reused only after its previous readers finish.
With symmetric user buffers, these buffers come from the NCCL-registered pool (Section~\ref{subsec:gtp-symmetric-buffers}).
At each training step, the new batch is copied into static input buffers before replay.
DDP's communication across replicas is also captured in the graph, as in eager execution: all-gathers of the updated weight shards overlap with the forward pass, and gradient reduce-scatters overlap with the backward pass.
The captured dependencies ensure that each updated shard arrives before GTP all-gathers it, and that GTP communication and the reduce-scatters finish before the optimizer runs.
The optimizer then runs eagerly or in its own graph.

\paragraph{Capturing dynamic MoE routing.}
Routing decisions and per-expert token counts can change between training steps, while graph replay requires stable buffer addresses.
HybridEP~\cite{2026hybridep} and the fused grouped-expert kernels preallocate a per-GPU token capacity shared across local experts, with saved activations held in paged storage within the same budget~\cite{2026mcoremoetraining}.
The routing decisions and token counts remain dynamic, but dispatch, expert computation, and activation storage use the provisioned capacity without host-side buffer sizing during replay (Figure~\ref{fig:gtp-full-iteration-graphs}(b)).
This trades capacity padding and storage headroom for graph-compatible execution.
If routing exceeds the preallocated token capacity or paged storage, MCore discards the attempt, deletes the graph, and reruns the same batch eagerly, with dropless dispatch and without paged storage; the next training step recaptures the graph.
The capacity bound concerns token storage; GTP continues to replay the recorded expert-weight prefetch order.

\subsection{AdamW and Muon on Sharded Weights}
\label{subsec:gtp-optimizers}

AdamW~\cite{2019adamw} updates each element independently, so it works directly on the stored shards.
Muon~\cite{2024muon} computes each update from a whole matrix, which no single GPU stores, so GTP must decide where that computation runs.
Section~\ref{subsec:gtp-eval-muon} measures the effect on training throughput.

\subsubsection{AdamW}
\label{subsubsec:gtp-adam}

MCore's distributed optimizer treats each stored shard as an ordinary parameter, and partitions its optimizer state across the $c/r$ GPUs that hold replicas of that shard.
Each GPU thus holds $\frac{1}{t \cdot r} \cdot \frac{r}{c} = \frac{1}{n}$ of each weight's optimizer state, even when $r < c$ and the weights themselves are divided across only $t \cdot r$ GPUs (Section~\ref{subsec:gtp-memory}).
A replicated weight (Section~\ref{par:gtp-selective-sharding}) has its optimizer state divided across $r$ times fewer GPUs.
The global gradient norm used for clipping counts each element once, excluding duplicates on other replicas and padding rows.
After each update, GTP re-quantizes each stored shard to its storage format (Section~\ref{subsec:gtp-low-precision}).
Section~\ref{subsec:gtp-eval-loss-curves} compares training and validation losses with and without GTP under AdamW.

\subsubsection{Muon}
\label{subsubsec:gtp-muon}

\begin{figure}[!htbp]
    \centering
    \includegraphics[width=\linewidth]{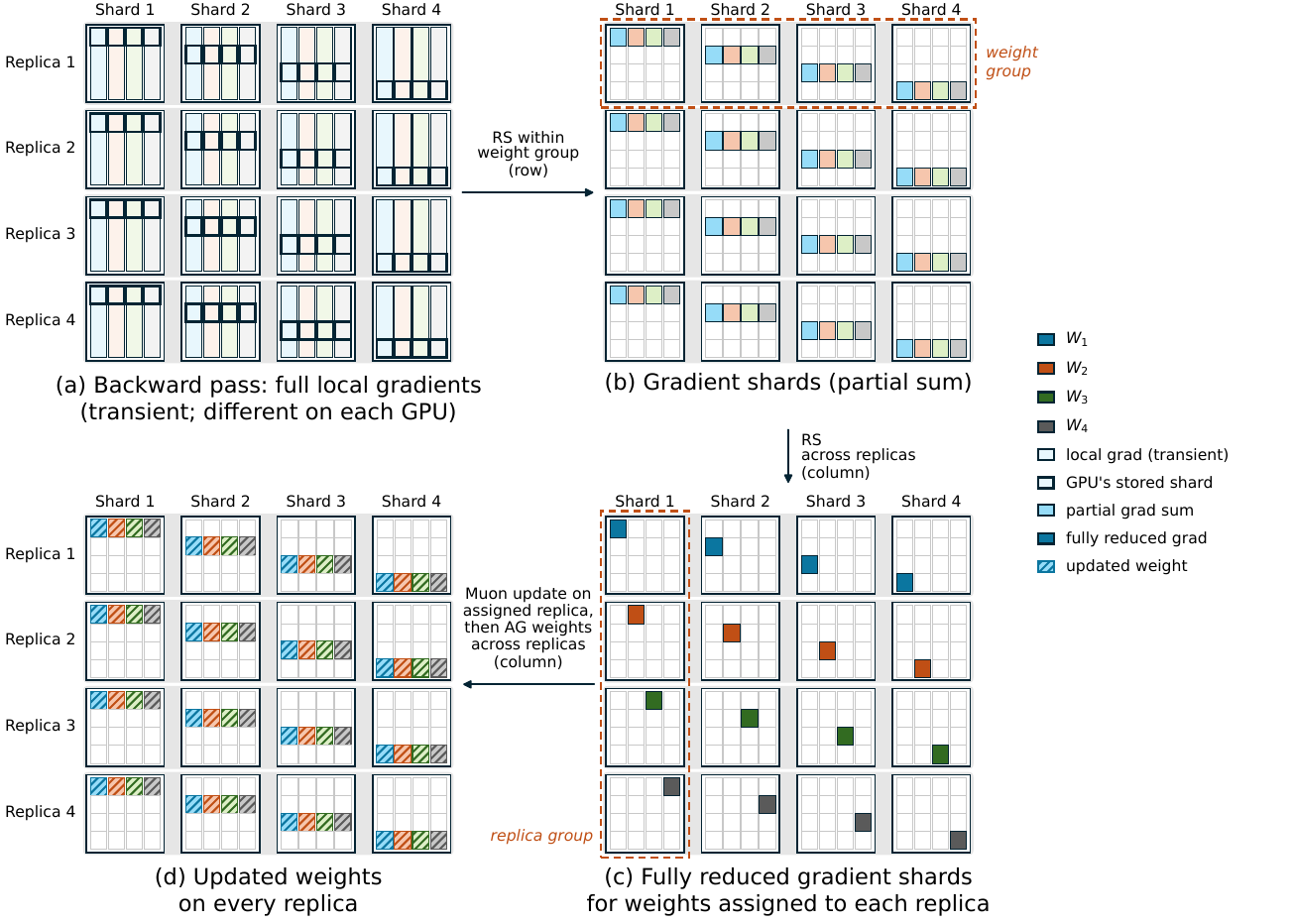}
    \caption{Hierarchical gradient reduction for Muon with the padded layout, which reduce-scatters gradient shards across replicas (16 GPUs, $t=1$, $r=4$).
    Each replica forms one weight group of four GPUs, and each of the four weights ($W_1$--$W_4$) has four replicas.
    Clockwise: (a) transient full local gradients; (b) partial sums sharded within weight groups; (c) fully reduced shards on assigned replicas; (d) updated weights all-gathered across replicas.
    In (a), bold outlines mark the stored shards.
    Colors identify weights; gradient values in (a) and (b) differ across replicas.
    Dashed outlines mark collective groups.}
    \label{fig:gtp-muon-copies}
\end{figure}

Unlike AdamW, Muon does not update each element independently: it orthogonalizes each weight's \emph{whole} momentum matrix, a running average of the weight's gradients, with a fixed number of Newton--Schulz (NS) iterations.
However, under GTP, each weight matrix and its gradient are split across $p = t \cdot r$ GPUs: $t$ TP partitions, each split across a weight group of $r$ GPUs.
Running Muon on sharded weights thus requires additional communication and a choice of which GPUs compute each matrix's update.

MCore's distributed optimizer shards optimizer state across the $c/r$ replicas of each weight: it assigns each matrix's update to one replica, the matrix's \emph{assigned replica}, whose $p$ GPUs each keep a $1/p$ shard of the matrix's momentum.
Gradients are first reduce-scattered within each weight group, leaving each GPU with its shard of every matrix's gradient, summed over only that weight group's tokens.
Gradient reduction across replicas then sums these partial shards over all replicas.
An all-reduce gives every GPU the fully reduced shards of every matrix, although each replica needs only those of the matrices it updates.
A reduce-scatter avoids this extra communication by giving each GPU the fully reduced values for only one slice of its gradient buffer.
It requires the gradient buffer to be sharded the same way as optimizer state: each GPU's slice must hold exactly the gradient shards of the matrices its replica updates.
The \emph{padded layout} ensures this by grouping each GPU's gradient shards by the replica that updates them and padding the groups to equal size, so it can use a reduce-scatter (Figure~\ref{fig:gtp-muon-copies}).
The \emph{compact layout}, used in our DSV4-Pro Muon evaluation, omits this padding and uses an all-reduce.

In both layouts, the assigned replica computes the update, and the updated weight shards are then all-gathered across replicas once per optimizer step (Figure~\ref{fig:gtp-muon-copies}(c--d)).
Within the assigned replica, each matrix's momentum is still split across $p$ GPUs, but NS operates on the whole matrix.
MCore supports three ways to run NS on these shards:

\begin{figure}[!tbp]
    \centering
    \includegraphics[width=\linewidth]{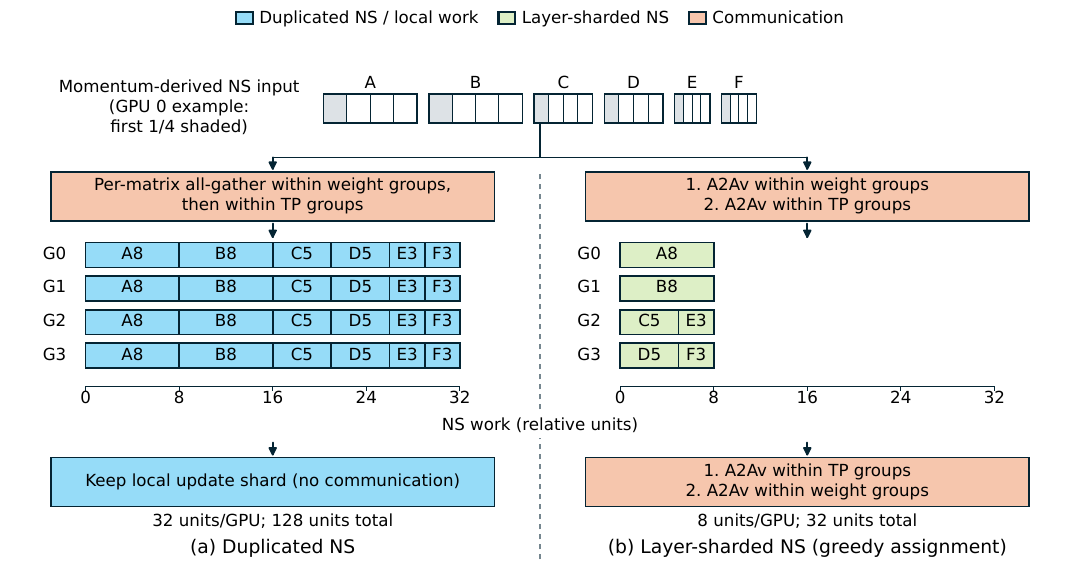}
    \caption{Layer sharding removes duplicated NS work.
    Illustrative GTP4 with TP2 ($p=4$, $t=2$, $r=2$): (a) per-matrix all-gathers let every GPU compute
    all six updates (32 units/GPU); (b) packed variable-size all-to-all (A2Av)
    routes inputs to greedily balanced homes, where each update runs once
    (8 units/GPU), then returns update shards. Tensor widths follow the
    illustrative NS-cost scale; shaded quarters show GPU 0's input shards.}
    \label{fig:gtp-muon}
\end{figure}

\begin{itemize}[leftmargin=*, itemsep=4pt, topsep=4pt]
\item \textbf{Duplicated NS} all-gathers the momentum and runs NS on each of these $p$ GPUs, reproducing unsharded Muon exactly.
Increasing $p$ does not reduce its per-GPU NS work (Figure~\ref{fig:gtp-muon}(a)).

\item \textbf{Distributed NS} keeps the momentum sharded and all-reduces a Gram matrix in every NS iteration.
Each GPU computes a partial Gram matrix from its shard, and the all-reduce sums them.
The Gram matrix is square in the unsharded dimension, so its all-reduce is smallest when that dimension is the smaller one.

\item \textbf{Layer-sharded NS} assigns each matrix an \emph{NS home}, one of the $p$ GPUs, using a greedy assignment that balances NS cost~\cite{1969graham}.
Packed variable-size all-to-alls (A2Av) send the momentum shards to the homes, which compute each update once and return the update shards (Figure~\ref{fig:gtp-muon}(b)), similar to owner-based updates in TorchTitan~\cite{2026torchtitanmuon} and FlexShard~\cite{2026flexshard}.
\end{itemize}
These three approaches trade off redundant computation (duplicated), communication in every NS iteration (distributed), and temporary memory and load balance (layer-sharded): each home holds whole matrices, and a matrix cannot be split across homes.
Duplicated NS serves as the numerical reference; distributed NS differs from it only by floating-point rounding, and layer-sharded NS matches it bitwise in our tests with default kernels.

\begin{figure}[!htbp]
    \centering
    \includegraphics[width=0.86\linewidth]{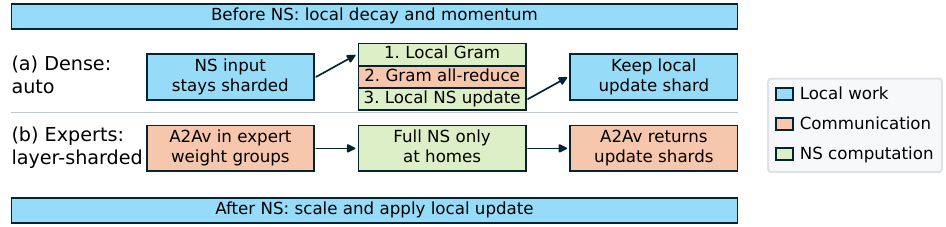}
    \caption{Hybrid Muon policies. (a) Distributed NS for a dense matrix, when the cost model selects it;
    steps 1--3 repeat in every NS iteration. (b) Layer-sharded experts at TP1.}
    \label{fig:gtp-muon-hybrid}
\end{figure}

In our runs, we choose how to run NS separately for dense and expert weights (\emph{hybrid Muon}\phantomsection\label{par:gtp-muon-hybrid-policy}; Figure~\ref{fig:gtp-muon-hybrid}).
Each replica is assigned only a few large dense matrices, which layer sharding cannot balance well across its $p$ GPUs, so MCore chooses between duplicated and distributed NS for each dense matrix with a cost model based on the matrix's shape, peak compute, and NVLink bandwidth.
Each replica is assigned many expert matrices of the same shape, which balance well, so expert weights use layer-sharded NS.
A few packed all-to-alls per optimizer step then replace the Gram all-reduces in every NS iteration, which would cross slower inter-domain links.

\subsection{Checkpointing}
\label{subsec:gtp-checkpointing}

GTP saves model and optimizer state through MCore's distributed checkpointing, which builds on PyTorch's distributed checkpoint (DCP) library~\cite{2026pytorchdcp}.
Each saved shard records its offset within the full weight, so a checkpoint can be loaded with different TP, GTP, EGTP, and data-parallel degrees.
Low-precision weights are saved in BF16.
Loading pads the logical rows for the new configuration (Section~\ref{subsec:gtp-padding}); a GPU whose stored shard falls entirely in the padding receives only zeros.

%% file: sections/5_evaluation.tex
\section{Evaluation}
\label{sec:evaluation}

We seek to answer the following main questions about GTP in this section:
\begin{itemize}[leftmargin=*, itemsep=2pt, topsep=4pt]
\item Does GTP match the loss curves of baseline training runs (Section~\ref{subsec:gtp-eval-loss-curves})?
\item How well does GTP scale to 12,288 GPUs (Section~\ref{subsec:gtp-eval-scaling})?
\item How does GTP compare with FSDP and 3D parallelism on throughput and memory at up to 1,024 GPUs (Sections~\ref{subsec:gtp-eval-fsdp2}--\ref{subsec:gtp-eval-3d})?
\item How do the design and implementation choices in Sections~\ref{sec:design} and~\ref{sec:implementation} affect throughput and memory (Sections~\ref{subsec:gtp-eval-long-context}--\ref{subsec:gtp-eval-padding})?
\end{itemize}

\subsection{Experimental Setup}
\label{subsec:gtp-eval-setup}

\paragraph{Hardware.} We use NVIDIA GB300 GPUs, four per node, each with approximately 298~GB of high-bandwidth memory (HBM).
The GPUs are organized into 72-GPU NVLink domains~\cite{2026gb300nvl72} with 900~GB/s of NVLink bandwidth per GPU, and nominal inter-domain network bandwidth of 100~GB/s per GPU.

\paragraph{Implementation and software.} GTP is integrated into MCore and TE (Section~\ref{sec:implementation}).
We use PyTorch 2.13, a TE 2.19 development build, CUDA 13.3, and NCCL~2.31.2.
We use Emerging Optimizers 0.3.1~\cite{2026emergingoptimizers,2026soapmuon} for an implementation of Muon.

\paragraph{Baselines.} We compare GTP with MFSDP~\cite{2026mfsdpbaselinebase} and PyTorch FSDP2 in TorchTitan~\cite{2026fsdp2schedsource,2026torchtitanschedsource}.
Both shard training state and re-materialize parameters around each module's execution.
We also compare with 3D parallelism (TP, PP, and DP, combined with EP for MoE layers) using a recipe adapted from Megatron Bridge~\cite{2026bridgedsv4prorecipe}.
Each throughput comparison matches the model configuration, GPU count, sequence length, global batch size, compute precision, and activation offloading and recomputation policy.
VPP denotes virtual (interleaved) pipeline parallelism~\cite{2021megatronscaling}.
A suffix gives a parallelism's degree; for example, GTP64 denotes 64-way dense parameter sharding and EGTP4 four-way expert parameter sharding.
Unspecified degrees are one.
MBS and GBS denote the microbatch and global batch sizes in sequences, and GA the number of gradient-accumulation microbatches per training step.
Unless otherwise specified, performance benchmarks use random initialization, AdamW, MXFP8 precision, and forced MoE load balancing.

\begin{table}[!htbp]
    \centering
    \small
    \setlength{\tabcolsep}{5pt}
    \begin{tabular}{@{}lcccccc@{}}
        \toprule
        \multirow{2}{*}{\textbf{Model}} & \textbf{Total} & \textbf{Active} & \textbf{Layers} & \textbf{Experts /} & \multirow{2}{*}{\textbf{\boldmath$h$ / $h'$}} & \textbf{Sequence} \\
        & \textbf{parameters} & \textbf{parameters} & \textbf{(full / proxy)} & \textbf{\boldmath top-$k$} & & \textbf{length} \\
        \toprule
        NT3-Nano~\cite{2025nemotron3nano} & 31.6B & 3.2B & 52 / -- & 128 / 6 & 2688 / 1856 & 8192 \\
        DSV3~\cite{2024deepseekv3} & 671B & 37B & 61 / 30 & 256 / 8 & 7168 / 2048 & 4096 \\
        DSV4-Pro~\cite{2026deepseekv4} & 1.6T & 49B & 61 / 30 & 384 / 6 & 7168 / 3072 & 4096 \\
        Internal-2T & 2T & 130B & 164 / 55 & 512 / 10 & 3072 / 10240 & 8192; $2^{16}$--$2^{20}$ \\
        \bottomrule
    \end{tabular}
    \caption{Benchmark model configurations.
    Parameter counts are for the full model; active parameters are those used per token.
    The hidden size is $h$, and each expert's intermediate size is $h'$ (Section~\ref{subsec:gtp-framework}).
    Proxies retain layer dimensions; -- denotes no proxy.
    Layers exclude multi-token prediction (MTP) layers~\cite{2024deepseekv3}: DSV3 and DSV4-Pro have one MTP layer; Internal-2T has two in the scale-out experiments and none in the proxy and long-context experiments.
    Sequence lengths are in tokens.}
    \label{tab:gtp-model-configs}
\end{table}

\paragraph{Workloads.} We evaluate GTP on DeepSeek-V3 (DSV3)~\cite{2024deepseekv3}, DeepSeek-V4-Pro (DSV4-Pro)~\cite{2026deepseekv4}, and Internal-2T, an internal 2T-parameter model.
Table~\ref{tab:gtp-model-configs} summarizes full-model configurations and proxy depths.
We abbreviate token and batch counts with $K=2^{10}$ and $M=2^{20}$; for example, 8K denotes 8,192 tokens.
To fit on fewer GPUs, some comparisons use proxy models with fewer layers but unchanged layer dimensions.
We check correctness on Nemotron-3-Nano (NT3-Nano)~\cite{2025nemotron3nano}.

\paragraph{Performance metrics.} We report \textbf{Tokens/s/GPU} and \textbf{Model TFLOP/s/GPU}~\cite{2026bridgeperformance}, which counts forward- and backward-pass FLOPs, excluding recomputation.
For comparisons across Megatron and TorchTitan, we use tokens/s/GPU because their native model-FLOP estimators differ, including causal-attention accounting~\cite{2026megatrondsv3flops,2026torchtitanattentionflops}.
Throughput is averaged over training steps after a warmup period.
Per-GPU memory denotes total device memory usage, including memory reserved by PyTorch's caching allocator but not currently allocated, in decimal GB; we report the maximum observed usage across the sampled GPUs.

\paragraph{Profiling metrics.} Exposed communication is the total time during which at least one communication kernel runs and no compute kernel runs on the profiled GPU.
We report it as a fraction of step time, except for backward-pass all-gathers, which we report as a fraction of the time during which any backward-pass all-gather runs.
Communication categories can overlap, so their exposed fractions do not sum to the total.

\subsection{Loss Curves}
\label{subsec:gtp-eval-loss-curves}

\begin{figure}[!htbp]
    \centering
    \includegraphics[width=\linewidth]{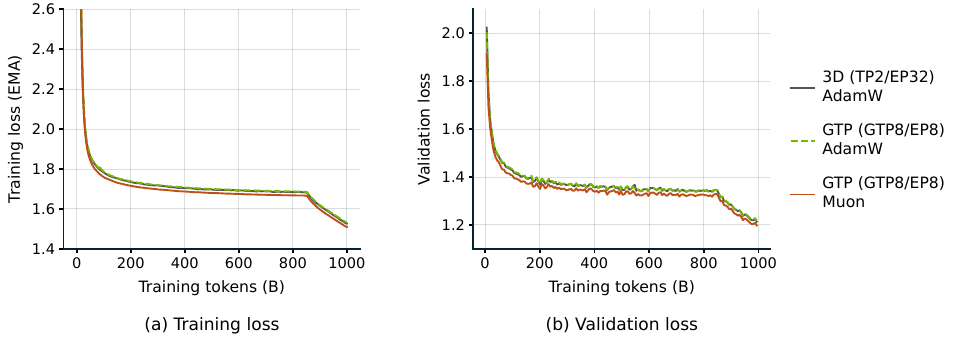}
    \caption{Nemotron-3-Nano loss curves over 1T tokens (8K sequences, GBS768).
    The AdamW runs compare 3D parallelism on 768 GPUs with GTP on 256 GPUs.
    Training loss is smoothed with an exponential moving average; validation loss is evaluated every 1,000 training steps.
    GTP uses dense parameter sharding (GTP8) without expert parameter sharding.
    The additional GTP--Muon curve uses a different optimizer and has no matched 3D-parallelism baseline.}
    \label{fig:nemotron3-nano-loss}
\end{figure}

Figure~\ref{fig:nemotron3-nano-loss} compares Nemotron-3-Nano training and validation losses with and without GTP.
The GTP AdamW run applies AdamW directly to the stored shards, as described in Section~\ref{subsubsec:gtp-adam}.
On Nemotron-3-Nano, trained for 158,900 training steps (roughly 1T tokens), GTP's mean absolute validation-loss difference from the baseline is 0.0022 nats, with a mean absolute relative difference of 0.16\%, over 158 matched validation checkpoints.
Both AdamW runs use MXFP8; the baseline uses TP2 and EP32 on 768 GPUs, and GTP uses GTP8 and EP8 on 256 GPUs.
The figure also includes GTP with Muon to demonstrate optimizer support; it has no matched 3D-parallelism baseline, so its lower loss should not be attributed to GTP.
We also obtained matching loss curves with EGTP on 100B- and 500B-parameter internal models.

\subsection{Throughput at Scale}
\label{subsec:gtp-eval-scaling}

\begin{table}[!htbp]
    \centering
    \small
    \setlength{\tabcolsep}{2pt}
    \begin{minipage}[t]{0.49\linewidth}
        \centering
        \textbf{(a) Weak scaling}\par
        \emph{One sequence per GPU}\par\smallskip
        \begin{tabular}{@{}ccccc@{}}
            \toprule
            \multirow{2}{*}{\textbf{GPUs}} & \multirow{2}{*}{\textbf{GBS}} & \textbf{TFLOP/s} & \textbf{Aggregate} & \textbf{Scaling} \\
            & & \textbf{per GPU} & \textbf{(EFLOP/s)} & \textbf{efficiency} \\
            \toprule
            4,096 & 4,096 & 1,079 & 4.42 & -- \\
            8,192 & 8,192 & 1,084 & 8.88 & 100.5\% \\
            12,288 & 12,288 & 1,060 & 13.03 & 98.2\% \\
            \bottomrule
        \end{tabular}
    \end{minipage}\hfill
    \begin{minipage}[t]{0.49\linewidth}
        \centering
        \textbf{(b) Strong scaling}\par
        \emph{Fixed GBS of 12,288 sequences}\par\smallskip
        \begin{tabular}{@{}ccccc@{}}
            \toprule
            \multirow{2}{*}{\textbf{GPUs}} & \multirow{2}{*}{\textbf{GBS}} & \textbf{TFLOP/s} & \textbf{Aggregate} & \textbf{Scaling} \\
            & & \textbf{per GPU} & \textbf{(EFLOP/s)} & \textbf{efficiency} \\
            \toprule
            6,144 & 12,288 & 1,156 & 7.10 & -- \\
            12,288 & 12,288 & 1,060 & 13.03 & 91.7\% \\
            \bottomrule
        \end{tabular}
    \end{minipage}
    \caption{Internal-2T training throughput at scale.
    GBS counts sequences.
    Per-GPU throughput is the median, over a run's logging intervals, of each 10-step interval's mean throughput.
    Aggregate throughput is per-GPU throughput times GPU count, in model EFLOP/s.}
    \label{tab:gtp-production-scaling}
\end{table}

With GTP, Internal-2T training reaches \textbf{13.0 model EFLOP/s} (1,060 TFLOP/s/GPU) on 12,288 GPUs with PP1, MBS1, and 8K sequences \emph{without forced load balancing} (Table~\ref{tab:gtp-production-scaling}).
These full-model runs use Muon (momentum 0.9, 16 NS iterations with Polar Express coefficients~\cite{2025polarexpress}) and eager execution.
Scale-out increases the number of data-parallel replicas while keeping GTP64, EGTP8, and EP64 fixed.
With one sequence per GPU (GA1), weak-scaling efficiency from 4,096 to 12,288 GPUs is \textbf{98.2\%}.
At a fixed GBS of 12,288 sequences, halving the GPU count from 12,288 to 6,144 (doubling GA to 2) increases per-GPU throughput by 9\%.

\subsection{Comparison with FSDP Implementations}
\label{subsec:gtp-eval-fsdp2}

We compare GTP with two FSDP implementations: MFSDP, weak-scaled from 256 to 1,024 GPUs on three models, and PyTorch FSDP2 in TorchTitan, in a matched comparison on the DSV3 proxy at 256 GPUs.
GTP achieves 1.05--1.55$\times$ MFSDP's throughput with 4--16\% lower per-GPU memory, and achieves 1.04$\times$ FSDP2's throughput with 16\% lower per-GPU memory.

\subsubsection{Weak Scaling against MFSDP}
\label{subsubsec:gtp-eval-mfsdp-scaling}

\begin{figure}[!htbp]
    \centering
    \includegraphics[width=\textwidth]{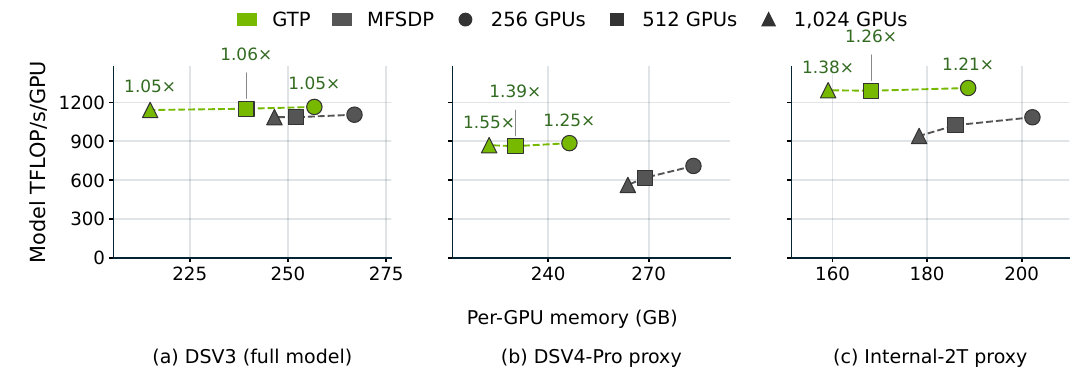}
    \caption{Weak-scaling throughput and memory for GTP vs. MFSDP (GA1): (a) full DSV3, (b) DSV4-Pro proxy, and (c) Internal-2T proxy.
    Throughput is in model TFLOP/s/GPU; labels show GTP's speedup over MFSDP at the same GPU count.
    Dashed lines connect each method's points in GPU-count order to show weak-scaling trends.
    MFSDP baselines are based on MCore 0.20.0 development code~\cite{2026mfsdpbaselinebase}, with model-specific extensions and fixes.
    }
    \label{fig:gtp-mfsdp-comparison}
\end{figure}

We weak-scale GTP and MFSDP from 256 to 1,024 GPUs on full DSV3 and the DSV4-Pro and Internal-2T proxies (Figure~\ref{fig:gtp-mfsdp-comparison}(a--c)).
DSV3 and DSV4-Pro use MBS2 and full-iteration CUDA graphs; the Internal-2T proxy uses MBS1 and eager execution.
All comparisons use EP64.
GTP shards dense weights across 64 GPUs within an NVLink domain; MFSDP shards parameters and gradients across 64 GPUs, and shards optimizer state across all GPUs.
GTP also uses EGTP4, whereas MFSDP does not shard expert weights beyond EP.

At 256 GPUs, GTP achieves \textbf{1.05$\times$}, \textbf{1.25$\times$}, and \textbf{1.21$\times$} MFSDP's throughput on full DSV3, the DSV4-Pro proxy, and the Internal-2T proxy, respectively (Figure~\ref{fig:gtp-mfsdp-comparison}(a--c)).
Across 256--1,024 GPUs, GTP's per-GPU memory is 4--13\%, 13--16\%, and 7--11\% lower than MFSDP's, respectively.

\paragraph{GTP's advantage over MFSDP.}
\phantomsection
\label{para:gtp-mfsdp-gain}
In the evaluated MFSDP configurations, backward hooks wait for all parameter all-gathers of a Transformer block (DSV3 and DSV4-Pro) or, for Internal-2T, of a pair of consecutive layers (a linear attention layer or a global softmax attention layer, and the MoE layer after it) before starting that block's backward computation.
Each GTP GEMM instead waits only for its own weights (Section~\ref{subsec:gtp-fsdp-problems}).
These MFSDP configurations do not shard expert weights beyond EP, so they never all-gather expert weights, and GFC1 never waits for GFC2's weights.
On DSV4-Pro at 256 GPUs, profiles show that 20\% of GTP's backward-pass all-gather time is exposed, versus 80\% for MFSDP, and that GTP's backward pass is 18.3\% shorter.
Gains are smaller on DSV3, which has more tokens per expert per microbatch under balanced routing than DSV4-Pro and Internal-2T (16,384, versus 8,192 and 10,240).
More tokens per expert provide more computation per all-gathered byte (Equation~\ref{eq:gtp-overlap-ratio}), which leaves less room for GTP's scheduling to improve overlap.

\paragraph{GTP's advantage grows with scale.}
\phantomsection
\label{para:gtp-weak-scaling}
From 256 to 1,024 GPUs, GTP's weak-scaling efficiency is \textbf{98.0\%, 98.2\%, and 98.7\%} on DSV3, DSV4-Pro, and Internal-2T, respectively.
MFSDP's is only 79.3\% on DSV4-Pro and 86.7\% on Internal-2T, so at 1,024 GPUs GTP achieves 1.55$\times$ and 1.38$\times$ MFSDP's throughput on these models.
On DSV4-Pro, both methods' compute time changes by less than 2\% from 256 to 1,024 GPUs, so MFSDP's step time grows from increased exposed communication time, mostly in the backward pass.
MFSDP's exposed backward-pass all-gather time is much larger at 1,024 GPUs than at 256 GPUs (1.26 vs. 0.34~s in profiled steps) and varies widely across training steps.
At 512 GPUs, this time ranges from 390 to 932~ms across steps, and nearly all of the difference is spent waiting for the last GPU in each group to start each all-gather.
GTP's exposed backward-pass all-gather time stays nearly flat from 256 to 1,024 GPUs.

\subsubsection{Matched Comparison with PyTorch FSDP2}
\label{subsubsec:gtp-eval-fsdp2-matched}

We compare GTP with PyTorch FSDP2 in TorchTitan~\cite{2026fsdp2schedsource,2026torchtitanschedsource} on the DSV3 proxy at 256 GPUs, because full DSV3 does not fit in the evaluated FSDP2 configuration.
Both use MBS2, GBS512, GA1, EP64, and 64-way dense sharding.
We use EGTP4 for GTP, and TorchTitan (with a local dense-mesh adapter) likewise shards expert weights four ways.
GTP and TorchTitan keep multi-head latent attention (MLA) down-projections replicated.
TorchTitan falls back to BF16 all-gathers for dense weights whose shards do not meet MXFP8 alignment; its expert all-gathers use MXFP8.

\begin{table}[!htbp]
    \centering
    \small
    \begin{tabular}{@{}lccc@{}}
        \toprule
        \multirow{2}{*}{\textbf{Method}} & \textbf{Tokens/s} & \textbf{Relative} & \textbf{Memory} \\
        & \textbf{per GPU} & \textbf{rate} & \textbf{(GB/GPU)} \\
        \toprule
        PyTorch FSDP2 (TorchTitan) & 8,503 & 1.00$\times$ & 219 \\
        GTP & 8,862 & 1.04$\times$ & 183 \\
        \bottomrule
    \end{tabular}
    \caption{Throughput and memory for the DSV3 proxy
    on 256 GPUs. Relative rates are normalized to FSDP2.}
    \label{tab:dsv3-matched-token-throughput}
\end{table}

\begin{figure}[!htbp]
    \centering
    \includegraphics[width=\linewidth]{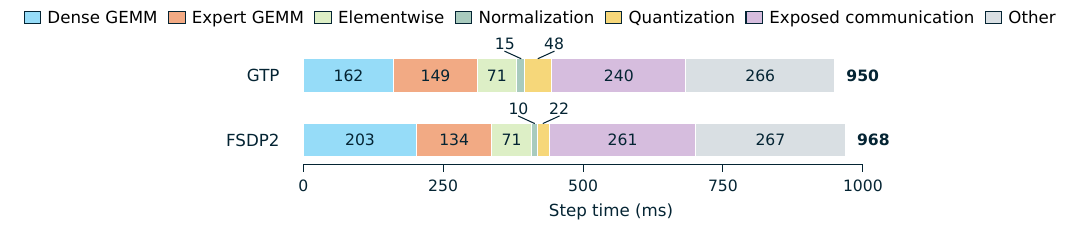}
    \caption{Step-time breakdown for the DSV3 proxy on 256 GPUs.
    Each interval of the step is assigned to one category, so the segments sum to the step time.
    Quantization counts standalone kernels; FSDP2's expert GEMMs also include fused EP token exchange and quantization.
    Elementwise covers the remaining pointwise and reduction kernels.
    Other includes attention, optimizer work, copies, and idle time.}
    \label{fig:dsv3-step-breakdown}
\end{figure}

\paragraph{Higher throughput and lower memory than FSDP2.}
GTP achieves 1.04$\times$ FSDP2's throughput (Table~\ref{tab:dsv3-matched-token-throughput}).
Full-step profiles show less time in dense GEMMs and exposed communication, but more in expert GEMMs and standalone quantization (Figure~\ref{fig:dsv3-step-breakdown}).
The quantization difference (48 vs.\ 22~ms) partly reflects fusion boundaries: TorchTitan's Dist-MoE backend~\cite{2026distmoe} includes EP exchange and input quantization inside expert GEMMs, whereas GTP executes HybridEP kernels~\cite{2026hybridep} and quantizers separately.
Normalization takes 15 vs.\ 10~ms, and the remaining elementwise work takes about 71~ms in both methods.

GTP uses 16\% less GPU memory than FSDP2 (183 vs.\ 219~GB per GPU).
GTP limits temporary storage through MXFP8 weight all-gathers and release of all-gather and reduce-scatter buffers after their last use; TorchTitan's BF16 all-gathers for unaligned dense shards need larger buffers.

\subsection{Comparison with 3D Parallelism}
\label{subsec:gtp-eval-3d}

\begin{figure}[!htbp]
    \centering
    \includegraphics[width=\linewidth]{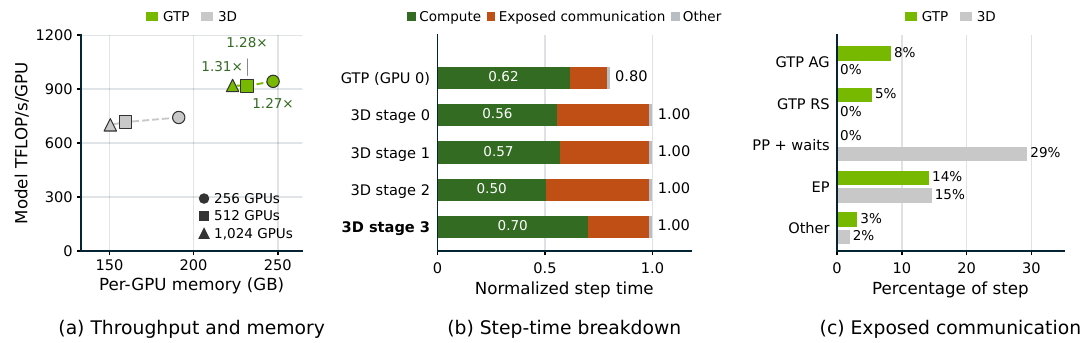}
    \caption{GTP vs. 3D parallelism on the DSV4-Pro proxy (GTP with MBS2; 3D parallelism with MBS1).
    (a) Weak scaling at 256--1,024 GPUs; labels show GTP's speedup.
    (b--c) Profiles from separate 256-GPU runs at GBS1K: (b) step time normalized to 3D stage 0; (c) exposed communication as a fraction of each method's step (GPU 0 for GTP; stage 0 for 3D parallelism).
    \textit{Other}: copies and idle time in (b); DP communication and remaining synchronization in (c); \textit{PP} includes peer waiting.}
    \label{fig:gtp-3d-comparison}
    \label{fig:dsv4-3d-step-breakdown}
    \label{fig:dsv4-3d-step-breakdown-appendix}
\end{figure}

We compare GTP and 3D parallelism on the DSV4-Pro proxy at 256--1,024 GB300 GPUs.
For weak scaling, we hold the work per GPU at four 4K sequences per training step, giving GBS1K at 256 GPUs.
3D parallelism uses TP1, PP4, VPP4, and EP64, with 8, 8, 7, and 7 layers per stage; GTP uses GTP64, EGTP4, and EP64.
Both use full-iteration CUDA graphs, offload core-attention activations to host memory, and selectively recompute MLA up-projections and manifold-constrained hyper-connections (mHC)~\cite{2025mhc}.

\paragraph{GTP and PP face different microbatch tradeoffs.}
We tune each method's MBS for throughput under the same hardware memory constraints.
Larger microbatches can improve GEMM efficiency for both methods and help GTP hide weight all-gathers.
At fixed GBS, however, they leave PP with fewer microbatches to fill the pipeline and increase activation memory.
At 256 GPUs and GBS1K:
\begin{itemize}[leftmargin=*, itemsep=2pt, topsep=4pt]
    \item \textbf{GTP saves memory at matched MBS.} GTP uses \textbf{11\% and 13\% less memory} than 3D parallelism at MBS1 and MBS2, respectively (169 vs.\ 191 and 247 vs.\ 285~GB/GPU).
    GTP does not retain activations for multiple in-flight microbatches, and it shards dense weights more finely than TP1 with PP4 does.
    \item \textbf{GTP trades higher memory use for higher throughput.} Increasing MBS from one to two raises GTP's throughput from 601 to 944 model TFLOP/s/GPU and memory from 169 to 247~GB/GPU.
    For 3D parallelism, the same change reduces throughput from 743 to 698 model TFLOP/s/GPU while raising memory from 191 to 285~GB/GPU.
\end{itemize}
For the scaling comparison, we use MBS2 for GTP and MBS1 for 3D parallelism, giving two microbatches per training step for GTP and 16 per pipeline for 3D parallelism.
With these choices, GTP achieves \textbf{1.27--1.31$\times$ the throughput of 3D parallelism} across 256--1,024 GPUs while using \textbf{29--48\% more memory} (Figure~\ref{fig:gtp-3d-comparison}(a)).

\paragraph{Stage imbalance slows 3D parallelism.}
To examine this advantage, we use profiles of both methods at 256 GPUs and GBS1K (Figure~\ref{fig:dsv4-3d-step-breakdown}(b--c)).
Per-GPU compute time is similar: about 2.3~s per step for GTP, versus a mean of 2.1~s across the pipeline stages in the 3D-parallelism baseline.
These stages are imbalanced, however: the last stage's compute time exceeds the stage mean by \textbf{20\%}, and the pipeline runs at the speed of its slowest stage.
The other stages spend the difference waiting, which appears as PP transfers and peer waiting: 29\% of stage 0's step, compared with an ideal balanced-stage bubble of 4.5\%~\cite{2021megatronscaling}.
GTP has no stages to balance, so its step is \textbf{20\% shorter}, even though its weight all-gathers and gradient reduce-scatters expose 8\% and 5\% of the step.

\begin{figure}[!htbp]
    \centering
    \includegraphics[width=0.54\linewidth]{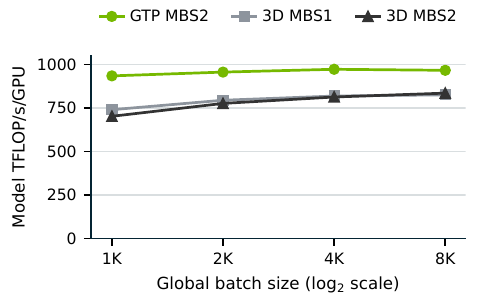}
    \caption{Throughput of the DSV4-Pro proxy on 256 GB300 GPUs with 4K sequences as GBS varies, for GTP with MBS2 and 3D parallelism with MBS1 and MBS2.}
    \label{fig:dsv4-gbs-sweep}
\end{figure}

\paragraph{Larger global batches narrow GTP's throughput advantage.}
On 256 GPUs with 4K sequences, increasing GBS from 1K to 8K improves the best throughput of 3D parallelism (across MBS1 and MBS2) by 13\%, versus 3\% for GTP, since larger batches amortize pipeline fill and drain over more microbatches (Figure~\ref{fig:dsv4-gbs-sweep}).
GTP's advantage narrows from \textbf{26\% to 16\%}, but it remains faster at every measured GBS.
The remaining gap reflects the stage imbalance described above.
Changing the MBS does not change this: with MBS2, 3D parallelism ranges from 5\% slower to 1\% faster than with MBS1 across GBS, and uses about 90~GB more memory per GPU.
These are separate runs from Figure~\ref{fig:gtp-3d-comparison}, so absolute numbers differ slightly.

\subsection{Long-Context Training}
\label{subsec:gtp-eval-long-context}
\label{subsec:gtp-eval-long-sequence}

To evaluate long-context training, we vary the sequence length of full Internal-2T from 8K to 1M tokens on 512 GPUs, keeping 4M tokens per training step (Table~\ref{tab:internal2t-long-context-configs}).
As sequence length grows, we increase the CP degree from 1 to 128 so that each GPU continues to process 8K tokens.
All configurations in the sweep use AdamW, GTP64, EGTP8, EP64, MBS1, and the same activation recomputation and offloading policy.

\begin{table}[!htbp]
    \centering
    \small
    \setlength{\tabcolsep}{5pt}
    \begin{tabular}{@{}cccccc@{}}
        \toprule
        \textbf{Sequence} & \multirow{2}{*}{\textbf{CP}} & \multirow{2}{*}{\textbf{GBS}} & \textbf{Step} & \textbf{Memory} & \textbf{Model} \\
        \textbf{length} & & & \textbf{time (s)} & \textbf{(GB/GPU)} & \textbf{TFLOP/s/GPU} \\
        \toprule
        8K & 1 & 512 & 5.1 & 272 & \textbf{1,258} \\
        64K & 8 & 64 & 7.0 & 280 & 1,127 \\
        256K & 32 & 16 & 12.4 & 290 & 1,018 \\
        1M & 128 & 4 & 35.8 & 283 & \textbf{883} \\
        \bottomrule
    \end{tabular}
    \caption{Long-context results at fixed 4M tokens per step on 512 GPUs, with forced balanced routing and eager execution.
    GBS counts sequences per step.
    }
    \label{tab:internal2t-long-context-configs}
\end{table}

\begin{figure}[!htbp]
    \centering
    \includegraphics[width=\linewidth]{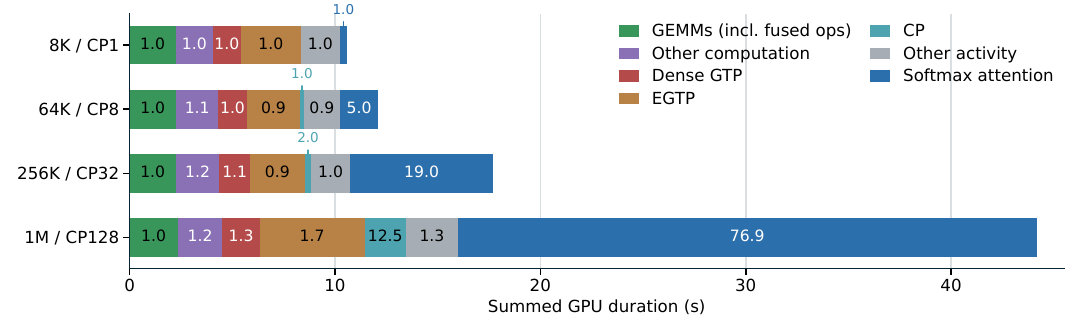}
    \caption{GPU-duration breakdown of the Internal-2T long-context sweep on 512 GPUs, \textbf{without subtracting overlap}.
    Widths sum kernel and copy durations; totals are not elapsed step times.
    Numeric labels show each component's normalized duration relative to its 8K value, except CP, which uses 64K as its baseline since 8K has no CP communication.
    \textit{Other computation} groups computation outside GEMMs and attention; \textit{other activity} groups HybridEP, gradient synchronization across replicas, remaining collectives, and copies.}
    \label{fig:internal2t-long-context-breakdown}
\end{figure}

With CP128, full Internal-2T runs training steps with 1M-token sequences at 283~GB per GPU.
From 8K to 1M, model throughput falls 30\%, from \textbf{1,258 to 883~TFLOP/s/GPU}, while step time rises from \textbf{5.1 to 35.8~s} (Table~\ref{tab:internal2t-long-context-configs}).

\paragraph{Softmax attention drives the growth in computation.}
At fixed $T$ tokens per step, softmax attention requires $O(TL)$ work for sequence length $L$.
Softmax attention duration grows $15.3\times$ from 64K to 1M for a $16\times$ increase in sequence length (Figure~\ref{fig:internal2t-long-context-breakdown}).
Across the sweep, GEMM duration varies by less than 5\%, while softmax attention, which runs in BF16, grows from 3\% to 64\% of summed GPU activity, reducing the share spent on faster MXFP8 GEMMs.

\paragraph{Outer CP adds inter-domain traffic.}
At 256K, each CP32 group fits within one NVLink domain.
At 1M, each sequence spans two 64-GPU weight groups, giving CP128 across two NVLink domains while dense GTP64 stays within each domain (Section~\ref{subsubsec:gtp-cp-folding}).
\begin{itemize}[leftmargin=*, itemsep=2pt, topsep=4pt]
\item \textbf{CP communication grows and rarely overlaps computation at 1M.}
From 64K with CP8 to 1M with CP128, summed CP duration grows $12.5\times$ (Figure~\ref{fig:internal2t-long-context-breakdown}), as exchanges of softmax-attention key-value pairs and of recurrent states from linear attention layers span more sequence partitions and cross the slower InfiniBand (IB) links between domains.
At 1M, 99.8\% of CP time has no concurrent computation, although 47.6\% overlaps other communication.

\item \textbf{Weight all-gathers take longer at 1M, partly from waiting on other GPUs.}
At 1M, summed dense GTP and EGTP durations reach $1.3\times$ and $1.7\times$ their 8K values (Figure~\ref{fig:internal2t-long-context-breakdown}), even though weight payloads and collective counts are unchanged.
Each EP64 group fills one NVLink domain, so each EGTP8 group spans all eight domains and therefore four CP128 groups; its all-gathers wait for the slowest of those groups and share IB links with CP.
The added EGTP time with no concurrent computation is about 1~s of the 35.8~s step.
\end{itemize}

\subsection{Per-GEMM Scheduling Ablations}
\label{subsec:gtp-eval-ablations}

\begin{figure}[!htbp]
    \centering
    \includegraphics[width=\linewidth]{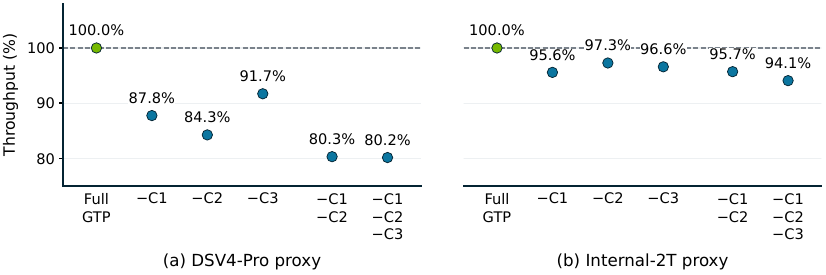}
    \caption{Scheduling ablations on (a) DSV4-Pro proxy and (b) Internal-2T proxy,
    both on 256 GPUs and normalized to each model's full GTP (100\%).
    Points show full GTP, independent removals of
    \hyperref[subsubsec:gtp-projection-chains]{C1},
    \hyperref[subsubsec:gtp-precise-waits]{C2}, and
    \hyperref[subsubsec:gtp-expert-coalescing]{C3},
    and cumulative removals of C1 and C2 and of all three.
    }
    \label{fig:gtp-component-ablation}
\end{figure}

We remove separate expert prefetch chains (\hyperref[subsubsec:gtp-projection-chains]{C1}), consumer-specific waits (\hyperref[subsubsec:gtp-precise-waits]{C2}), and expert-weight coalescing (\hyperref[subsubsec:gtp-expert-coalescing]{C3}), individually and cumulatively, on two benchmark models, holding all other settings fixed (Figure~\ref{fig:gtp-component-ablation}(a--b)).
Both use GTP64, EGTP4, EP64, and GA1; DSV4-Pro uses MBS2 and GBS512, and Internal-2T uses MBS1 and GBS256.
DSV4-Pro uses full-iteration CUDA graphs; Internal-2T uses eager execution.

Without C1, GFC1 and GFC2 share one expert prefetch chain.
Without C2, a GEMM that consumes a prefetched weight also waits for all work already queued on that weight's all-gather stream, including prefetches for other GEMMs.
Without C3, each local expert weight is all-gathered separately rather than coalesced with the other weights feeding the same grouped GEMM.

\paragraph{Each mechanism improves throughput in both workloads.}
Independently removing C1, C2, and C3 reduces throughput by \textbf{12.2\%, 15.7\%, and 8.3\%} on DSV4-Pro and \textbf{4.4\%, 2.7\%, and 3.4\%} on Internal-2T, respectively; the largest individual loss comes from C2 on DSV4-Pro and C1 on Internal-2T.

\paragraph{Gains are larger on DSV4-Pro.}
Removing all three mechanisms reduces throughput by \textbf{19.8\% on DSV4-Pro} and \textbf{5.9\% on Internal-2T}.
Under balanced routing, Internal-2T processes 25\% more tokens per expert per microbatch (10,240 versus 8,192), providing more arithmetic per all-gathered weight (Equation~\ref{eq:gtp-overlap-ratio}).
Its smaller scheduling loss is consistent with more computation covering communication, although the models and execution modes both differ.

\paragraph{Gains are not additive.}
Individual losses sum to 36.2\% on DSV4-Pro and 10.5\% on Internal-2T, but removing all three costs only 19.8\% and 5.9\%, respectively.
Once C1 and C2 are removed, also removing C3 lowers throughput from 80.3\% to 80.2\% of full GTP on DSV4-Pro, a difference too small to resolve without repeated runs, and from 95.7\% to 94.1\% on Internal-2T.

\subsection{Muon on Sharded Weights}
\label{subsec:gtp-eval-muon}

\begin{figure}[!htbp]
    \centering
    \includegraphics[width=0.7\linewidth]{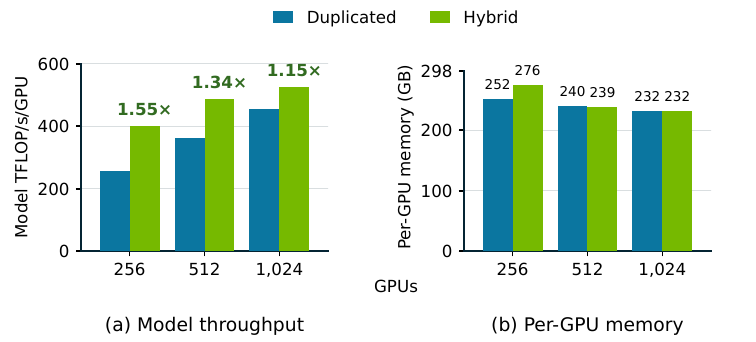}
    \caption{Duplicated and hybrid Muon on the DSV4-Pro proxy (8K; GBS equals GPU count; one run per point).
    (a) Model throughput, with hybrid token-throughput speedups labeled.
    (b) Peak per-GPU memory.}
    \label{fig:gtp-dsv4-muon}
\end{figure}

We compare hybrid Muon (Section~\ref{subsubsec:gtp-muon}) with duplicated NS on the DSV4-Pro proxy at 256, 512, and 1,024 GPUs (Figure~\ref{fig:gtp-dsv4-muon}).
All three scales use 8K sequences, GTP64, EGTP4, EP64, MBS1, and GA1.
Both Muon modes use momentum 0.95 and ten NS iterations with DeepSeek-V4's coefficients~\cite{2026deepseekv4}.
CUDA graphs are disabled.

\paragraph{Hybrid Muon improves throughput at every scale, with diminishing gains.}
Hybrid Muon improves token throughput by \textbf{$1.55\times$ at 256 GPUs, $1.34\times$ at 512 GPUs, and $1.15\times$ at 1,024 GPUs} as GPU count and GBS scale together (Figure~\ref{fig:gtp-dsv4-muon}(a)).
These runs measure the combined configuration and do not separate the dense and expert contributions.

\paragraph{More replicas reduce per-GPU NS work.}
At 256, 512, and 1,024 GPUs, each dense weight has 4, 8, and 16 replicas, and each expert weight has 1, 2, and 4, while tokens per GPU and sharding degrees stay fixed.
Each replica updates a smaller share of the matrices (Section~\ref{subsubsec:gtp-muon}), so each GPU performs fewer NS updates in both modes.
Less optimizer work shortens the step while forward- and backward-pass work stays roughly unchanged, so per-GPU throughput rises.
Less NS work per GPU also narrows hybrid Muon's relative advantage (Figure~\ref{fig:gtp-dsv4-muon}(a)).

\paragraph{Layer sharding's extra memory shrinks with scale.}
At 256 GPUs, hybrid Muon uses about 23~GB more device memory than duplicated NS (Figure~\ref{fig:gtp-dsv4-muon}(b)).
Each NS home temporarily holds whole FP32 matrices and exchange buffers (Section~\ref{subsubsec:gtp-muon}).
As GPU count grows, the expert matrices assigned to each four-GPU weight group for expert weights fall from 372 to 186 to 93, shrinking these buffers.
At 512 and 1,024 GPUs, device memory is comparable across the two modes.

\subsection{Symmetric User Buffers}
\label{subsec:gtp-eval-ub}

\begin{figure}[!htbp]
    \centering
    \includegraphics[width=0.9\linewidth]{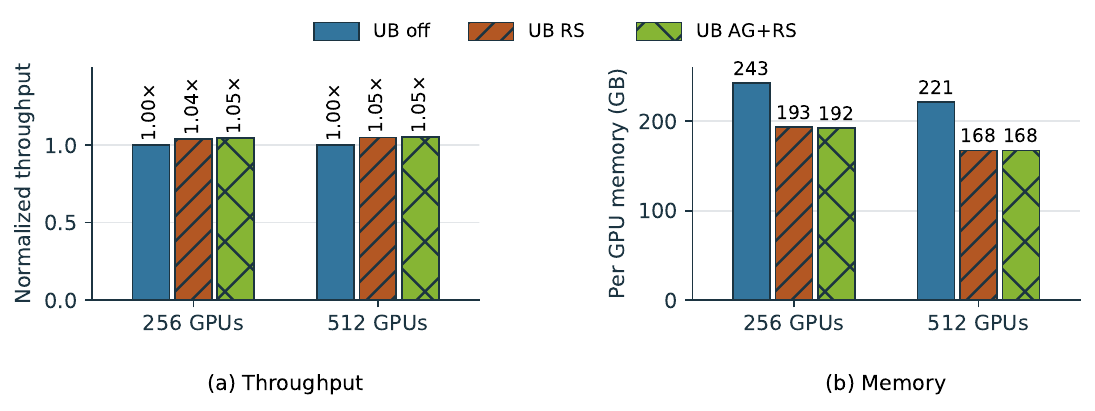}
    \caption{Dense user-buffer modes on the Internal-2T proxy at 256 and 512 GPUs: (a) throughput normalized to UB off at each GPU count; (b) per-GPU memory (GB).}
    \label{fig:nt-ub-scaling}
\end{figure}

We compare dense GTP with UB off, UB RS, and UB AG+RS (Section~\ref{subsec:gtp-symmetric-buffers}) on the Internal-2T proxy at 256 and 512 GPUs (Figure~\ref{fig:nt-ub-scaling}).
Expert-weight reduce-scatters keep the all-to-all path, and DDP's gradient collectives do not use UB.
All other settings, including eager execution, are fixed across modes.

\paragraph{Higher throughput with lower memory use.}
At 256 GPUs, UB RS and UB AG+RS improve throughput over UB off by 4.0\% and 4.4\% and reduce per-GPU memory by 20\% and 21\%, respectively.
At 512 GPUs, both modes improve throughput by about 5\% and reduce per-GPU memory by about 24\%.
UB RS captures most of the benefit at both scales, with little additional gain from registering all-gather buffers.
The memory savings come from avoiding the all-to-all path's receive buffers (Section~\ref{subsec:gtp-symmetric-buffers}).

\subsection{Sharding Granularity and Padding Tradeoffs}
\label{subsec:gtp-eval-padding}

Sharding a weight more finely shrinks each GPU's stored shard, but MXFP8 and NVFP4 require each shard's row count to be a multiple of their block size, so the added padding can shrink or even eliminate these savings (Section~\ref{subsec:gtp-padding}).
Table~\ref{tab:gtp-padding-overhead} quantifies the added weight elements relative to each GEMM's logical size at TP1.
These percentages describe weight storage and collective payloads, excluding scales, optimizer state, and temporary buffers.

\begin{table}[!htbp]
    \centering
    \small
    \setlength{\tabcolsep}{3pt}
    \begin{tabular}{@{}lcccccc@{}}
        \toprule
        \multirow{2}{*}{\textbf{GEMM}} & \multirow{2}{*}{\textbf{Logical shape}} & \multicolumn{3}{c}{\textbf{MXFP8}} & \multicolumn{2}{c}{\textbf{NVFP4}} \\
        \cmidrule(lr){3-5}\cmidrule(lr){6-7}
        & & \textbf{GTP32} & \textbf{GTP64} & \textbf{GTP128} & \textbf{GTP64} & \textbf{GTP128} \\
        \toprule
        \multicolumn{7}{l}{\textit{DSV3}} \\
        Attention output projection & $7{,}168\times16{,}384$ & 0.0\% & 14.3\% & 14.3\% & 0.0\% & 14.3\% \\
        Shared-expert FC2 & $7{,}168\times2{,}048$ & 0.0\% & 14.3\% & 14.3\% & 0.0\% & 14.3\% \\
        Q down-projection (if sharded) & $1{,}536\times7{,}168$ & 33.3\% & 33.3\% & 166.7\% & 33.3\% & 33.3\% \\
        \midrule
        \multicolumn{7}{l}{\textit{DSV4-Pro}} \\
        Attention output projection & $7{,}168\times16{,}384$ & 0.0\% & 14.3\% & 14.3\% & 0.0\% & 14.3\% \\
        Shared-expert FC2 & $7{,}168\times3{,}072$ & 0.0\% & 14.3\% & 14.3\% & 0.0\% & 14.3\% \\
        KV projection (if sharded) & $512\times7{,}168$ & 100.0\% & 300.0\% & 700.0\% & 100.0\% & 300.0\% \\
        \bottomrule
    \end{tabular}
    \caption{Calculated padding overhead for DSV3 and DSV4-Pro weights at TP1.
    Shared experts use dense GTP.
    Our GTP benchmarks keep DSV3's MLA down-projections and DSV4-Pro's KV weight replicated;
    rows marked ``if sharded'' show counterfactual costs.
    We do not run GTP128 or NVFP4.}
    \label{tab:gtp-padding-overhead}
\end{table}

\paragraph{Selective sharding avoids disproportionate padding.}
With MXFP8 and GTP64, attention output and shared-expert FC2 GEMMs incur \textbf{14.3\%} padding in both models.
Sharding DSV3's Q down-projection and DSV4-Pro's KV projection would incur \textbf{33.3\%} and \textbf{300\%} padding, respectively.
Whether to shard a weight should depend on its shape.

\paragraph{Padding can eliminate further memory savings for individual weights.}
Under MXFP8, moving either model's attention output projection from GTP32 to GTP64 reduces stored elements per GPU by 43\%, while increasing the padded payload by 14.3\%.
If DSV3's Q down-projection were sharded, GTP64 and GTP128 would both store 32 aligned rows per GPU: padding would rise from 33.3\% to 166.7\%, doubling the collective payload without further storage savings.
NVFP4 would eliminate padding for the attention output projections at GTP64, but would leave 33.3\% for the Q down-projection if sharded.

\paragraph{Padding increases storage and communication costs.}
GTP removes padded rows before GEMMs, so GEMM shapes and FLOPs are unchanged.
Padding enlarges stored shards and collective payloads; its throughput impact can depend on memory pressure and communication contention as well as exposed transfers.

\paragraph{Smaller expert sharding degrees avoid row padding.}
At TP1, routed FC1 and FC2 weights in DSV3 and DSV4-Pro require \textbf{zero MXFP8 row padding at EGTP2, EGTP4, and EGTP8}: their output dimensions are divisible by $32\times8=256$ rows.
Configuring dense GTP and EGTP independently permits wide dense sharding without imposing the same alignment cost on experts.

%% file: sections/6_parallelism_tradeoffs.tex
\section{Lessons for Configuring GTP}
\label{sec:design-space}

GTP's generality comes with many configuration choices, and the best choice depends on the model, training hyperparameters such as batch size, and the cluster.
This section briefly discusses how these choices should be made.

\subsection{Hiding Parameter All-Gathers}
\label{subsec:gtp-eval-overlap}

Parameter sharding trades stored-weight memory for all-gather traffic.
Overlapping the all-gathers with computation can partially hide this cost.
The freed memory also allows larger microbatches, which provide more computation to hide the all-gathers.

\paragraph{Tokens per expert set the overlap budget.}
Equation~\ref{eq:gtp-overlap-ratio} explains why tokens per expert matter when configuring GTP: more tokens provide more computation to cover each weight all-gather.
Under balanced routing, tokens per expert per microbatch range from 8,192 for DSV4-Pro to 16,384 for DSV3; Section~\ref{subsubsec:gtp-eval-mfsdp-scaling} relates these differences to the observed scheduling gains.
Larger microbatches, longer local sequences, and larger EP degrees raise $s_e$; gradient accumulation does not, because it repeats both the computation and the all-gathers.
Raising the EGTP degree $r_e$ moves the all-gathered fraction $1-1/r_e$ only slowly toward 1; the main cost of a larger $r_e$ is that all-gathers spanning NVLink domains see a much lower $B_r^e$.
For dense weights, our within-domain placement keeps all-gathers on NVLink (Section~\ref{subsec:gtp-weight-data-views}), reducing all-gather time relative to cross-domain placement.

\paragraph{Scheduling limits overlap.}
A high ratio in Equation~\ref{eq:gtp-overlap-ratio} is an opportunity, not a guarantee: the schedule must deliver each weight before its consumer runs.
GFC1 and GFC2 need their weights at different times; with consumer-specific waits (C2), a GEMM stalls only if its own weights are not ready when its other inputs are (Section~\ref{subsec:gtp-scheduling}).
On DSV4-Pro, disabling consumer-specific waits (C2) alone reduces throughput to 84.3\% of full GTP at unchanged token count and all-gather volume (Section~\ref{subsec:gtp-eval-ablations}).

\paragraph{Memory limits how far $s$ can grow.}
Increasing $s$ raises activation memory.
Increasing EP can raise tokens per expert without increasing input tokens per GPU, but changes dispatch buffers and local-expert counts; keeping EP within an NVLink domain also caps its degree at the domain size.
Prefetched weights also occupy memory, limiting the lookahead (Section~\ref{subsec:gtp-fsdp-problems}).
Recomputation or offloading frees memory at added computation or transfer cost.

\subsection{Choosing Parallelism Degrees and Placement}
\label{subsec:gtp-placement-guidance}

Configuring GTP means choosing the TP, re-materialization ($r$ for dense weights, $r_e$ for expert weights), and CP degrees, and placing each group of GPUs relative to NVLink domains.
Each choice trades communication against memory; the balance depends on tokens per GPU and layer shapes.

\paragraph{TP degree.}
TP's activation communication grows with $t$, whereas parameter communication shrinks; Equation~\ref{eq:gtp-tp-balance} gives the TP degree that minimizes the combined communication time assuming a bandwidth-only model.
Small TP degrees suit many tokens per GPU, and larger degrees suit few tokens, wide layers, or fast TP links; for MoEs, $h'$ is each expert's intermediate size, not the number of experts.

\paragraph{Re-materialization degrees.}
Larger re-materialization degrees shrink each GPU's stored weights and gradients to $1/(t \cdot r)$ of each dense weight and $1/r_e$ of each expert weight, but each all-gather then spans more GPUs.
The $t \cdot r$ GPUs that together hold one replica of each dense weight should stay within an NVLink domain, which keeps per-microbatch weight all-gathers on fast links.
Expert weights can instead use a larger $r_e$ that spans domains (Section~\ref{subsec:gtp-cross-domain-experts}): their all-gathers are prefetched on separate chains (C1) and do not grow with tokens, provided the ratio in Equation~\ref{eq:gtp-overlap-ratio} stays high at the lower cross-domain bandwidth.
Small weights can remain replicated when sharding saves little memory or introduces disproportionate padding (Sections~\ref{par:gtp-selective-sharding} and~\ref{subsec:gtp-eval-padding}).

\paragraph{CP degree.}
At fixed sequence length and MBS, increasing CP reduces tokens per GPU and their activation storage.
With GPU count and TP and PP degrees also fixed, it leaves fewer independent batch partitions.
Keeping GBS fixed then requires more gradient-accumulation microbatches.
These extra microbatches repeat weight all-gathers, while fewer tokens per GEMM can reduce compute efficiency.
The attention backend must also communicate over a larger CP group.
For long sequences, increasing CP in proportion to sequence length keeps tokens per GPU and the number of microbatches fixed, as in our long-context experiments (Section~\ref{subsec:gtp-eval-long-context}).

\paragraph{EP and CP placement.}
EP should stay within an NVLink domain (e.g., EP64 in a 72-GPU domain~\cite{2026gb300nvl72}), because dispatch and combine feed dependent computation and grow with local tokens.
CP should also stay within an NVLink domain when activation memory permits.
When CP must span domains, each weight group for dense weights should still stay within an NVLink domain (Section~\ref{subsubsec:gtp-batch-sequence}), and CP then shares the inter-domain network with EGTP (Section~\ref{subsec:gtp-eval-long-context}).

\subsection{Tradeoffs with Pipeline Parallelism}
\label{subsec:gtp-pp-tradeoffs}

\paragraph{Memory.}
PP and GTP both reduce each GPU's parameter and optimizer-state footprint in proportion to their degrees.
In MoE models, most parameters are expert weights, so the relevant comparison is between the PP and EGTP degrees.
GTP also reduces activation memory in proportion to the number of GPUs it spans, but it adds a fixed buffer overhead for re-materialized weights and their gradients.
PP introduces a memory asymmetry: although each stage stores activations only for its own layers, the first stage retains them for as many in-flight microbatches as there are stages, so its activation memory does not shrink.

\paragraph{Throughput.}
PP's main throughput overhead is pipeline fill and drain, which is amortized over the microbatches in a training step and therefore shrinks as the number of microbatches grows.
Virtual pipelining shortens fill and drain by assigning each GPU several smaller model chunks, at the cost of more point-to-point communication~\cite{2021megatronscaling}.
PP also needs computation balanced across stages, which is harder for hybrid models whose layer types differ in cost, and harder still with virtual pipelining, since each chunk contains fewer layers to balance.
GTP's main throughput overhead is its weight all-gathers, which are expensive but can be hidden when each microbatch has enough computation and communication is carefully scheduled (Equation~\ref{eq:gtp-overlap-ratio}).
In our measurements, larger global batches narrowed GTP's advantage over 3D parallelism as fill and drain were amortized, but stage imbalance kept 3D parallelism slower at every measured batch size (Section~\ref{subsec:gtp-eval-3d}).

\paragraph{Choosing between GTP and PP.}
GTP alone is preferable when activation memory dominates or when each microbatch has enough computation to hide its all-gathers.
Adding PP helps when the global batch is large enough to amortize fill and drain and the stages can be balanced, or when GTP's all-gathers cannot be hidden.

%% file: sections/7_related_work.tex
\section{Related Work}
\label{sec:related}

\paragraph{Unified sharding abstractions.}
GSPMD~\cite{2021gspmd} and Mesh-TensorFlow~\cite{2018meshtensorflow} express data, tensor, and other forms of parallelism as sharding annotations on tensor dimensions over a device mesh, and GSPMD's compiler propagates these annotations to every operator.
Alpa~\cite{2022alpa} searches over such intra-operator plans jointly with pipeline stages.
GTP's view of DP, TP, SP, CP, and FSDP as partitioning and re-materialization choices is in the same spirit.
We focus on what this view implies for communication scheduling: we treat parameter re-materialization as a preprocessing step on one input of a GEMM, including GEMMs inside fused operators, and make it efficient at scale in an eager framework.

\paragraph{FSDP implementations.}
\phantomsection
\label{subsec:fsdp2}
ZeRO-3~\cite{2020zero} and PyTorch FSDP~\cite{2023pytorchfsdp} shard parameters, gradients, and optimizer state, and re-materialize parameters per module.
MFSDP~\cite{2026mfsdpbaselinebase}, PyTorch FSDP2~\cite{2026fsdp2api}, veScale-FSDP~\cite{2026vescalefsdp}, and FlexShard~\cite{2026flexshard} expose controls for prefetching, bucketing, and grouping parameters, and DeepSpeed ZeRO-3 prefetches parameters in the module order recorded during a traced training step~\cite{2026deepspeedzero3source}.
These systems grew out of data parallelism and share a unit-based scheduling model.
They differ in how units are formed and in the prefetching, bucketing, and retention controls they expose.
Section~\ref{subsec:gtp-fsdp-problems} discusses the costs of this model for fused MoE layers.
Finer units, such as separate MFSDP units for FC1 and FC2, can address some of these costs but must be configured per model.
We evaluate GTP against MFSDP and PyTorch FSDP2 in Section~\ref{subsec:gtp-eval-fsdp2}.

veScale-FSDP and FlexShard are two other recent FSDP implementations, which focus on how parameters are distributed across GPUs.
veScale-FSDP uses RaggedShard layout planning and DBuffer views to group communication, align storage, and reduce copies~\cite{2026vescalefsdp}.
FlexShard aligns storage with optimizer ownership to avoid redistribution for Muon updates, and supports block-aligned FP8 layouts and configurable communication groups~\cite{2026flexshard}.
Its owner assignment balances whole matrices by parameter count, whereas GTP's layer-sharded NS balances them by NS cost.

MiCS~\cite{2022mics}, ZeRO++~\cite{2024zeroplusplus}, and HSDP~\cite{2023pytorchfsdp} confine parameter all-gathers to subgroups on fast links.
GTP's weight groups similarly keep weight all-gathers on fast links.
GTP also partitions data independently of the weight groups, so weight all-gathers do not need to be performed over full CP groups, and a long sequence can span several weight groups (Section~\ref{subsec:gtp-weight-data-views}).

\paragraph{3D parallelism.}
3D parallelism, which combines TP, PP, and DP, has been widely used to train large models~\cite{2019gpipe,2019pipedream,2021megatronscaling}, but its PP component has three weaknesses.
Pipeline fill and drain leave GPUs idle unless each training step has many microbatches.
Stages must be balanced, or faster stages wait for slower ones.
Each stage also holds activations for several in-flight microbatches.
Prior work reduces these costs: PipeDream uses profile-guided partitioning to balance pipeline stages~\cite[Section~3.1]{2019pipedream}, and Alpa jointly searches stage partitions and device assignments for heterogeneous layers~\cite[Sections~5 and~8.1]{2022alpa}.
Megatron-LM's virtual pipeline parallelism interleaves model chunks to reduce ideal pipeline bubbles at the cost of increased inter-stage communication~\cite[Section~2.2.2]{2021megatronscaling}.
GTP composes with PP; configurations without PP avoid stage imbalance and pipeline bubbles, but need enough memory to hold re-materialized weights and enough computation to hide the communication cost of re-materialization (Section~\ref{subsec:gtp-pp-tradeoffs}).

\paragraph{Overlapping TP communication.}
The activations communicated in TP are the output of the preceding computation, so overlapping this communication requires splitting the GEMM and the communication into smaller pieces (Section~\ref{subsec:gtp-framework}).
Wang et al.\ decompose collectives together with their dependent computation~\cite{2023overlapdecomposition}, T3 triggers communication from the producing GEMM with hardware support~\cite{2024t3}, FLUX fuses tiled communication into GEMM kernels~\cite{2024flux}, and Domino slices inputs and weights into independent pieces~\cite{2024domino}.
GTP's parameter all-gathers, in contrast, do not depend on preceding computation, so they can be issued ahead of the GEMMs that consume them without splitting any computation; these techniques still apply to GTP's activation communication when $t > 1$.

\paragraph{Parallelizing MoE layers.}
GShard~\cite{2021gshard}, DeepSpeed-MoE~\cite{2022deepspeedmoe}, Tutel~\cite{2023tutel}, and MegaBlocks~\cite{2023megablocks} accelerate MoE layers with techniques such as expert parallelism, runtime adaptation to changing expert workloads, and specialized kernels.
Our implementation builds on MoE Parallel Folding in MCore, which decouples the parallelism mappings of attention and MoE layers~\cite{2025moeparallelfolding}; GTP extends it by letting dense and expert weights use different re-materialization degrees (GTP and EGTP).
MegaScale-MoE customizes parallelism per module and overlaps MoE communication in production~\cite{2025megascalemoe}, and COMET overlaps EP communication with expert computation at fine granularity~\cite{2025comet}.
These systems make token exchange and expert computation more efficient; GTP composes with EP and with these optimizations.

%% file: sections/8_conclusion.tex
\section{Conclusion}
\label{sec:conclusion}

In this paper, we showed that, for linear layers, five commonly used parallelisms (DP, TP, SP, CP, and FSDP) are special cases of a new framework we call generalized tensor parallelism (GTP), in which parameters and activations are partitioned and either can be re-materialized.
Treating parameter re-materialization as a preprocessing step on one input of a GEMM, rather than a step before an entire Transformer block, lets us schedule all-gathers per GEMM.
With our implementation in MCore and TE, an internal 2T-parameter MoE model reaches 13.0 model EFLOP/s on 12,288 GB300 GPUs.
On 256--1,024 GPUs, our implementation achieves 1.05--1.55$\times$ MFSDP's throughput across three MoE models and 1.27--1.31$\times$ the throughput of a tuned 3D-parallel recipe on the DSV4-Pro proxy.

%% file: sections/9_acknowledgments.tex
\section*{Acknowledgments}

We thank Xiaowei~Ren, Yashaswi~Karnati, Peter~Dykas, Zhongbo~Zhu, Zhengyang~Wang, Hongxiao~Bai, Yuzhong~Wang, Sangkug~Lym, and Michael~Andersch for their contributions to this work.
We also thank colleagues not listed individually, and the open-source community for co-developed features, feedback, and continued engagement.